# Raman scattering in current carrying molecular junctions
A preliminary account

Michael Galperin,[1,*] Mark A. Ratner,[2] and Abraham Nitzan[2,3]

[1] Department of Chemistry and Biochemistry, University of California San Diego, La Jolla, CA 92093-0340, USA
[2] Department of Chemistry and Materials Research Center, Northwestern University, Evanston, IL 60208, USA
[3] School of Chemistry, The Sackler Faculty of Science, Tel Aviv University, Tel Aviv 69978, Israel

## Abstract

This is a preliminary acount of a theory for Raman scattering by current-carrying molecular junctions. The approach combines a non-equilibrium Green's function (NEGF) description of the non-equilibrium junction with a generalized scattering theory formulation for evaluating the light scattering signal. This generalizes our previous study (Phys. Rev. Lett. **95**, 206802 (2005); J. Chem. Phys. **124**, 234709 (2006)) of junction spectroscopy by including molecular vibrations and developing machinery for calculation of state-to-state (Raman scattering) fluxes within the NEGF formalism. For large enough voltage bias we find that the light scattering signal contains, in addition to the normal signal associated with the molecular ground electronic state, also a contribution from the inverse process originated from the excited molecular state as well as an interference component. The effect of coupling to the electrodes and of the imposed bias on the total Raman scattering as well as its components are discussed. Our result reduces to the standard expression for Raman scattering in the isolated molecule case, i.e. in the absence of coupling to the electrodes. The theory is used to discuss the charge transfer contribution to surface enhanced Raman scattering for molecules adsorbed on metal surfaces and its manifestation in the biased junction.

[*] Previous address: Theoretical Division and Center for Integrated Nanotechnologies (CINT), Los Alamos National Laboratory, Los Alamos, NM 87545, USA

## 1. Introduction

Surface enhanced Raman and resonance Raman spectroscopies (SERS and SERRS)[1, 2] have become important diagnostic tools for many science applications. Early observations of an apparent enhancement of the Raman signal of up to a few orders of magnitudes on rough surfaces and on particles of noble metals like silver, gold, and copper were explained by a combination of local electromagnetic field enhancement associated with surface plasmon excitations in structures of suitable size-range in such metals[3] and resonance scattering associated with charge transfer between the chemisorbed molecule and the metal substrate.[4-6] The behavior of local electromagnetic fields at metal and dielectric interfaces is a long studied problem, and the classical electromagnetic theory of SERS is by now reasonably well understood[7] although ongoing work requires detailed calculations on particular surface structures. On the other hand, the nature of the charge transfer contribution is still under discussion. Experimental indications of such contribution to SERS are mainly based on the different scattering behavior sometimes observed for molecules in the first adsorbate layer, and on observations of peaks in the SERS intensities of molecules adsorbed on electrode surfaces measured against electrode potential, whose positions shifts linearly with incident light frequency.[8] A theoretical treatment pertaining to the latter observations, in particular the shape and peak positions of SERS/voltage spectra, was given by Lombardi and coworkers[9-11] who have cast a model used earlier by Persson[6] in the framework of Herzberg-Teller-based Albrecht theory[12] of Raman scattering. Lombardi et al.[9-11] have attempted to explain the apparent discrepancy between the resonance nature of the charge transfer contribution to SERS and the lack of pronounced overtone peaks in the scattering signal by invoking the Herzberg-Teller intensity borrowing concept as used by Albrecht,[12] however this was done by assuming that terms in the Raman intensity in which such resonance structure appears can be disregarded.

An important development in the field was the observation of single-molecule SERS and SERRS[13-15] which led to the observation that much of the observed average SERS is associated with molecules adsorbed at particular "hot spots" where the enhancement was found to reach up to fourteen orders of magnitude. Indeed, studies of the electromagnetic field distribution in illuminated metal structures reveal the existence of spots with particularly strong field enhancement, e.g. positions located between two or more small metal particles.[15-17] Charge transfer between molecule and metal was suggested as a mechanism of blinking observed in the Raman signal from such hot spots.[18, 19] Of particular interest to our discussion are molecules adsorbed at contacts between metal electrodes, so called molecular junctions, whose electrical transport properties are under intensive studies.[20, 21] Some structures of this type, e.g. those based on junctions that comprise two gold spheres connected by a single molecule,[22] are similar to structures used as models for Raman hot spots. The possibility of monitoring Raman scattering and other optical processes together with electrical transport in such molecular junctions is of primary importance, both because of the interesting science of optical response of non-equilibrium molecular systems and because a successful accomplishment of this goal will establish SERS and SERRS as diagnostic tools for non-equilibrium systems while providing much needed tools to the field of molecular electronics. Indeed, this issue has already been discussed,[23] and preliminary experimental results in this direction have started to appear.[24-26]

In this paper we make a first step in the theoretical analysis of such systems by generalizing our recent treatment of optical response of molecular junctions to enable the description of Raman scattering. We avoid a detailed description of the electromagnetic enhancement by focusing on the molecular response to the *local* electromagnetic field, assuming the latter to be independent of the imposed potential bias. Our goal is to develop a theoretical approach capable of describing Raman scattering from biased, current carrying molecular junctions, where optical response can be used as a probe of the system's non-equilibrium state.

The structure of this paper is as follows: In section 2 we introduce the model that we use to describe Raman scattering in a molecular junction. Section 3 describes our theoretical approach to Raman scattering from molecular junctions. Two mechanisms for Raman scattering are considered. In one, the process is assumed to be dominated by the molecule-radiative field coupling. The other results from contributions from direct light induced charge transfer between the molecule and the metal substrate. In section 4 we present and discuss our numerical results. Section 5 concludes.

## 2. Model

We employ a generalization of the model used previously by Galperin and Nitzan,[27, 28] which comprises a molecule coupled to two metal electrodes (L and R, also referred to as source and drain) each in its own equilibrium. The molecule is represented by its highest occupied and lowest unoccupied molecular orbitals, HOMO and LUMO, respectively, that are used to describe the ground $(1,0)$ and lowest excited $(0,1)$ molecular states, as well as positive $(0,0)$ and negative $(1,1)$ ion states. Here the molecular states $(n_h, n_\ell)$ are represented by populations $n_h$ and $n_\ell$ of the HOMO and LUMO, respectively. The electrons on the molecule interact with the molecular vibrations, with electron-hole excitations in the leads $L$ and $R$, and with the radiation field. The latter is represented by photon modes of frequencies $\nu_\alpha$, whose polarization degrees of freedom are disregarded for simplicity. Also for simplicity we represent the molecular vibrations by a single harmonic oscillator which is in turn coupled to a thermal bath represented by continuum of such oscillators. In the linear response regime for the molecule-radiation field interaction it is sufficient to consider zero and single occupation of the radiation field modes. The steady state of the radiation field is accordingly described by one singly occupied mode of frequency $\nu_i$ (referred to as the pumping mode) with all other modes at zero occupancy. The observable of interest is the constant population flux from this pumping mode to another mode of frequency $\nu_f$. The existence of a continuum of radiation field modes is manifested by the usual radiative broadening of the excited molecular state. The Hamiltonian of the system reads (here and below we put $e=1$, $\hbar=1$, and $k_B=1$ for the electron charge and the Planck and Boltzmann constants, respectively)

$$\hat{H} = \hat{H}_0 + \hat{V}^{(e-v)} + \hat{V}^{(et)} + \hat{V}^{(\upsilon-b)} + \hat{V}^{(e-h)} + \hat{V}^{(e-p)} \tag{1}$$

where $\hat{H}_0$ includes additively all the subsystems Hamiltonians, while the $\hat{V}$ terms describe interactions between them. Here $(e-\upsilon)$ denotes interaction between the tunneling electron and the molecular vibration, $(et)$ – the coupling associated with electron transfer between molecule and leads, $(\upsilon-b)$ -- coupling between the molecular vibration and the thermal bath, $(e-h)$ stands for interaction between the molecular excitation and electron-hole excitations in the leads, and $(e-p)$ denotes the coupling of such molecular excitation and the radiation field. The explicit expressions for these terms are

$$\begin{aligned}\hat{H}_0 \;&= \sum_{m=1,2} \varepsilon_m \hat{d}_m^\dagger \hat{d}_m + \omega_\upsilon \hat{b}_\upsilon^\dagger \hat{b}_\upsilon + \sum_{k\in L,R} \varepsilon_k \hat{c}_k^\dagger \hat{c}_k \\ &+ \sum_\beta \omega_\beta \hat{b}_\beta^\dagger \hat{b}_\beta + \sum_{\alpha\in\{i,\{f\}\}} \nu_\alpha \hat{a}_\alpha^\dagger \hat{a}_\alpha\end{aligned} \tag{2}$$

$$\hat{V}^{(e-\upsilon)} = \sum_{m=1,2} V_m^{(e-\upsilon)} \hat{Q}_\upsilon \hat{d}_m^\dagger \hat{d}_m \tag{3}$$

$$\hat{V}^{(et)} = \sum_{K=L,R}\sum_{k\in K;m}\left(V_{km}^{(et)}\hat{c}_k^\dagger \hat{d}_m + V_{mk}^{(et)}\hat{d}_m^\dagger \hat{c}_k\right) \tag{4}$$

$$\hat{V}^{(\upsilon-b)} = \sum_\beta U_\beta^{(\upsilon-b)}\hat{Q}_\upsilon \hat{Q}_\beta \tag{5}$$

$$\hat{V}^{(e-h)} = \sum_{k_1\neq k_2}\left(V_{k_1k_2}^{(e-h)}\hat{D}^\dagger \hat{c}_{k_1}^\dagger \hat{c}_{k_2} + V_{k_2k_1}^{(e-h)}\hat{c}_{k_2}^\dagger \hat{c}_{k_1}\hat{D}\right) \tag{6}$$

$$\hat{V}^{(e-p)} = \hat{V}_M^{(e-p)} + \hat{V}_{CT}^{(e-p)} \tag{7a}$$

$$\hat{V}_M^{(e-p)} = \sum_{\alpha\in\{i,\{f\}\}}\left(U_\alpha^{(e-p)}\hat{D}^\dagger \hat{a}_\alpha + U_\alpha^{(e-p)*}\hat{a}_\alpha^\dagger \hat{D}\right) \tag{7b}$$

$$\hat{V}_{CT}^{(e-p)} = \sum_{\alpha\in\{i,\{f\}\}}\sum_{k\in\{L,R\}}\sum_{m=1,2}\left[V_{km,\alpha}^{(e-p)}\hat{D}_{km} + V_{mk,\alpha}^{(e-p)}\hat{D}_{mk}\right]\left(\hat{a}_\alpha + \hat{a}_\alpha^\dagger\right) \tag{7c}$$

where $\hat{d}_m^\dagger$ ($\hat{d}_m$) and $\hat{c}_k^\dagger$ ($\hat{c}_k$) create (annihilate) an electron in the molecular state $m$ and in the lead state $k$ of energies $\varepsilon_m$ and $\varepsilon_k$, respectively. $\hat{b}_\upsilon^\dagger$ ($\hat{b}_\upsilon$) and $\hat{b}_\beta^\dagger$ ($\hat{b}_\beta$) create (annihilate) vibrational quanta in the molecular mode, $\upsilon$, and the thermal bath mode, $\beta$, respectively. $\hat{a}_\alpha^\dagger$ ($\hat{a}_\alpha$) stands for creation (annihilation) operators of the radiation field quanta. $\omega$ and $\nu$ denote frequencies of phonon modes and of radiation field (photon) modes, respectively. Also

$$\hat{Q}_j \equiv \hat{b}_j + \hat{b}_j^\dagger \qquad j=\upsilon,\beta \tag{8}$$

are displacement operators for the molecular ($\upsilon$) and thermal bath ($\beta$) vibrations respectively and (for future reference)

$$\hat{P}_j \equiv -i\left(\hat{b}_j - \hat{b}_j^\dagger\right) \qquad j=\upsilon,\beta \tag{9}$$

are the corresponding momentum operators. Finally,

$$\hat{D} \equiv \hat{d}_1^\dagger \hat{d}_2; \qquad \hat{D}^\dagger \equiv \hat{d}_2^\dagger \hat{d}_1 \tag{10a}$$

are annihilation and creation operators for the molecular excitation (referred to below as molecular polarization operators), and similarly

$$\hat{D}_{mk} \equiv \hat{d}_m^\dagger \hat{c}_k; \quad \hat{D}_{km} \equiv \hat{c}_k^\dagger \hat{d}_m. \tag{10b}$$

The term $\hat{V}_M^{(e-p)}$, Eq. (7b), represents coupling of the radiation field to the transition between the ground and excited molecular states, while $\hat{V}_{CT}^{(e-p)}$, Eq. (7c), accounts for metal-molecule charge transfer optical transitions.[29] In our treatment below we will use zero and one photon occupation of the relevant radiation field modes. Therefore the coupling amplitudes $U_\alpha^{(e-p)}$ and $V_{km,\alpha}^{(e-p)}$ should reflect the intensity of the local electromagnetic field in the junction, including field enhancement effects associated with plasmon excitations in the leads. They depend on the photon frequency $\nu_\alpha$ through the standard factor $\sqrt{\nu_\alpha}$ that enters into the radiative coupling operator, as well as via this plasmonic response. Because photon frequencies relevant to our discussion span the relatively narrow range between the incoming and the outgoing radiation we will sometimes disregard this dependence as detailed below.

It should be noted that, while the vibronic coupling (3), where different electronic states are characterized by parallel-shifted harmonic nuclear potential surfaces, is common in molecular spectroscopy, it corresponds to standard treatments of Raman scattering only for near resonance processes. Far from resonance, the scattering amplitude between states $|g\upsilon_i\rangle$ and $|g\upsilon_f\rangle$ (where $\upsilon_i$ and $\upsilon_f$ are vibrational numbers associated with the ground electronic state $g$) is evaluated under the usual perturbative treatment that invokes approximations such as

$$\sum_{\upsilon_x} \frac{\langle g\upsilon_f | U_\alpha^{(e-p)} \hat{X} | x\upsilon_x \rangle \langle x\upsilon_x | X^\dagger U_\alpha^{(e-p)*} | g\upsilon_i \rangle}{\nu_i - \Delta E - \omega_v (\upsilon_x - \upsilon_i)} \cong \frac{\left\langle \upsilon_f \left| \left| \left(U_\alpha^{(e-p)}\right)_{gx} \right|^2 \right| \upsilon_i \right\rangle}{\nu_i - \Delta E} \tag{11}$$

Here $\left(U_\alpha^{(e-p)}\right)_{gx}$ is essentially the $gx$ element of the electronic dipole moment operator between electronic states $g$ and $x$, $\Delta E$ is their energy separation and $\hat{X}$ is the nuclear shift operator defined below. The resulting Raman contribution comes from the nuclear coordinate(s) dependence of $U_\alpha^{(e-p)}$ which is disregarded in our treatment. The present theory is therefore mostly suitable for resonance Raman processes while results for non-resonance situations described below should be regarded as qualitative.

Canonical (small polaron or Lang-Firsov) transformation[30-32] is next employed to eliminate electron-molecular vibration coupling (for detailed discussion see Ref. [33]), leading to the transformed Hamiltonian

$$\hat{\bar{H}} = \hat{\bar{H}}_0 + \hat{\bar{V}}^{(et)} + \hat{\bar{V}}^{(\upsilon-b)} + \hat{\bar{V}}^{(e-h)} + \hat{\bar{V}}^{(e-p)} \tag{12}$$

Explicit expressions for the right-hand-side terms in (12) are the same as in (2)-(7) with electron creation (annihilation) operators in the molecular subspace dressed by molecular vibration shift operators $\hat{X}_m$

$$\hat{X}_m \equiv \exp\left[i\lambda_m \hat{P}_\upsilon\right] \quad \lambda_m \equiv \frac{V_m^{(e-\upsilon)}}{\omega_\upsilon} \quad (m = 1,2) \tag{13}$$

so that

$$\hat{d}_m \rightarrow \hat{d}_m \hat{X}_m, \qquad m = 1,2 \tag{14}$$

$$\hat{D} \rightarrow \hat{D}\hat{X}\,;\;\; \hat{D}_{mk} \rightarrow \hat{D}_{mk}\hat{X}_m^{\dagger}\;;\;\; \hat{D}_{km} \rightarrow \hat{D}_{km}\hat{X}_m, \tag{15}$$

$$\hat{X} \equiv \hat{X}_1^{\dagger}\hat{X}_2 = \exp\left[i(\lambda_2 - \lambda_1)\hat{P}_{\upsilon}\right] \equiv \exp\left[i\lambda\hat{P}_{\upsilon}\right] \tag{16}$$

This transformation also shifts the molecular electronic orbital energies (polaronic shift) according to

$$\bar{\varepsilon}_m = \varepsilon_m - \lambda_m V_m^{(e-\upsilon)} \quad (m = 1,2) \tag{17}$$

Below we assume that this shift is taken into account and will drop the bar above $\varepsilon_m$.

The Hamiltonian (12) is the starting point of our treatment. We will also use another decomposition of this Hamiltonian

$$\hat{\bar{H}} \equiv \hat{\bar{H}}'_0 + \hat{\bar{V}}^{(e-p)} \tag{18}$$

where

$$\hat{\bar{H}}'_0 = \hat{\bar{H}}_0 + \hat{\bar{V}}^{(et)} + \hat{V}^{(\upsilon-b)} + \hat{\bar{V}}^{(e-h)} \tag{19}$$

is the Hamiltonian for the pure transport problem without coupling to the radiation field.

In the next section we advance a nonequilibrium Green's function-based formalism for describing resonance Raman scattering in non-equilibrium molecular junctions. Assuming that the molecular resonance energy $\Delta E$ and the charge transfer resonances (essentially the energy differences between the electronic chemical potentials in the metals and the molecular HOMO or LUMO) are well separated from each other, we treat their contributions separately. Below we refer by Model M to the system described by the Hamiltonian (2)-(7) with only the molecular radiative term (7 b). Model CT refers to the same Hamiltonian with only the charge transfer radiative interaction (7c).

## 3. Method

We focus first on model M, where the only contribution to the interaction (7) arises from the term (7b). With the goal of describing Raman spectroscopy in non-equilibrium junctions we consider the general non-equilibrium Green's function (NEGF)-based expression for the photon flux. The derivation essentially follows the standard consideration for electronic current in junctions[34, 35]; the difference is that in the present case Bose statistics has to be used for the carriers (photons). An equivalent expression for the thermal (phonon) flux has been previously derived by us[36] and others.[37-40] A potential complication in the present situation is that the molecular excitation operator $\hat{D}$ is not a true Bose operator. This however does not change the final result for the photon flux from mode $\alpha$ into the system, which is given by (see Appendix A for derivation)

$$\begin{aligned} J_\alpha(t) &\equiv -\frac{d}{dt} < \hat{a}_\alpha^\dagger(t)\hat{a}_\alpha(t) > \\ &= -\int_{-\infty}^{t} dt' \Big[ \Pi_\alpha^<(t,t')\mathcal{G}^>(t',t) + \mathcal{G}^>(t,t')\Pi_\alpha^<(t',t) \\ &\qquad - \Pi_\alpha^>(t,t')\mathcal{G}^<(t',t) - \mathcal{G}^<(t,t')\Pi_\alpha^>(t',t) \Big] \end{aligned} \tag{20}$$

where

$$\Pi_\alpha^{>,<}(t,t') = |U_\alpha^{(e-p)}|^2 F_\alpha^{>,<}(t,t') \tag{21}$$

is the greater (lesser) projection of the system self-energy (SE) due to coupling to the radiation field mode $\alpha$, $F_\alpha^{>,<}$ is the greater (lesser) projection of the free photon GF in mode $\alpha$

$$F_\alpha(\tau,\tau') \equiv -i < T_c \hat{a}_\alpha(\tau)\hat{a}_\alpha^\dagger(\tau') > \tag{22}$$

and $\mathcal{G}^{>,<}$ is the greater (lesser) projection of the molecular polarization GF (dressed by molecular vibration shift operator)

$$\mathcal{G}(\tau,\tau') \equiv -i < T_c \hat{D}(\tau)\hat{X}(\tau)\hat{D}^\dagger(\tau')\hat{X}^\dagger(\tau') > \tag{23}$$

Here $T_c$ is the time ordering operator on the Keldysh contour.[35, 41] Here and below we use $t$ to indicate real time variables, while $\tau$ is reserved for time variables on the Keldysh contour. At steady-state (20) simplifies to

$$J_\alpha = -\int_{-\infty}^{\infty} d(t-t') \Big[ \Pi_\alpha^<(t'-t)\mathcal{G}^>(t-t') - \Pi_\alpha^>(t'-t)\mathcal{G}^<(t-t') \Big] \tag{24}$$

where the first and second terms on the r.h.s. correspond to incoming and outgoing photon fluxes in the mode $\alpha$.

While the non-equilibrium state of the junction is best described within the NEGF formalism, a technical difficulty in using it in the present context stems from the fact that the Raman process under discussion is a scattering process, associated with well defined initial and final states, which is naturally described by scattering theory. Here we handle this situation by considering the photon flux through the molecule between two ”photon reservoirs”: One associated with the incoming radiation, in which only the incident mode $i$ is populated, and the other, associated with the outgoing radiation, where all modes are vacant. Using NEGF methodology to evaluate the steady state flux into the outgoing mode $f$ under these conditions yields results akin to scattering theory.

To achieve this goal consider first the outgoing photon flux from the system into a particular mode $f$, given by the second term in (24)

$$J_f = \int_{-\infty}^{+\infty} d(t-t')\Pi_f^{>}(t'-t)\mathcal{G}^{<}(t-t') \tag{25}$$

To relate this flux to the incident radiation field, the source term, i.e. the incoming flux taken into account in $\mathcal{G}^{<}$, should be related to the laser pumping mode $i$. Strictly speaking the GF $\mathcal{G}^{<}$ (lesser projection of Eq. (23)) is a two-particle GF in the electron subspace dressed by the many-particle shift operators $\hat{X}$. This would imply attempting to solve the Bethe-Salpeter equation (here complicated by the presence of the phonon shift operators, which make Wick's theorem inapplicable) with a kernel that describes a tunneling electron interacting with the pumping laser mode $i$. The SE notion is applicable in this case only approximately.[42] A simple workaround can be achieved by restricting consideration to the case of weak fields, hence taking into account only the lowest (second order) interaction with the mode $i$ on the Keldysh contour. Eq.(25) then leads to (see Appendix B)

$$\begin{aligned} J_{i\to f} = \int_{-\infty}^{+\infty} d(t-t') \Big[ \int_c d\tau_1 \int_c d\tau_2 \Pi_f^{>}(t'-t)\Pi_i(\tau_1,\tau_2) \\ \times < T_c \hat{D}^\dagger(t')\hat{X}^\dagger(t')\hat{D}(t)\hat{X}(t)\hat{D}^\dagger(\tau_1)\hat{X}^\dagger(\tau_1)\hat{D}(\tau_2)\hat{X}(\tau_2) > \Big] \end{aligned} \tag{26}$$

Note that in Eq. (26) $t$ and $t'$ are defined on the real time axis while $\tau_1$ and $\tau_2$ are defined on the Keldysh contour. Also note that although $t$ and $t'$ appear explicitly in (26), the integral inside the square brackets in this equation depends only on $t-t'$ as already implied by Eq. (25). Next, in Eq. (26) a projection of the variables $\tau_1$ and $\tau_2$ onto the real time axis has to be done. In doing so we again focus on the physics of interest -- the photon scattering process $i \to f$. Since the radiation mode $i$ is the source of the photon flux we disregard terms containing the outgoing photon self energy terms $\Pi_i^{>}(t_1-t_2)$ associated with this mode, keeping only projections containing the incoming self-energy $\Pi_i^{<}(t_1-t_2)$. Furthermore we keep only terms corresponding to rates, i.e. those where interaction with external field connects the upper and lower branches of the contour.[43] This leads to (for further details see Appendix B)

$$J_{i\to f} = J_{i\to f}^{(nR)} + J_{i\to f}^{(iR)} + J_{i\to f}^{(intR)} \tag{27}$$

where

$$J_{i\to f}^{(nR)} = |U_i|^2 |U_f|^2 \int_{-\infty}^{+\infty} d(t-t') \int_{-\infty}^{t} dt_1 \int_{-\infty}^{t'} dt_2 \, e^{-i\nu_i (t_1 - t_2)} e^{i\nu_f (t-t')} \\ < \hat{X}(t_2) \hat{X}^\dagger(t') \hat{X}(t) \hat{X}^\dagger(t_1) >< \hat{D}(t_2) \hat{D}^\dagger(t') \hat{D}(t) \hat{D}^\dagger(t_1) > \tag{28}$$

$$J_{i\to f}^{(iR)} = |U_i|^2 |U_f|^2 \int_{-\infty}^{+\infty} d(t-t') \int_{t}^{+\infty} dt_1 \int_{t'}^{+\infty} dt_2 \, e^{-i\nu_i (t_1 - t_2)} e^{i\nu_f (t-t')} \\ < \hat{X}^\dagger(t') \hat{X}(t_2) \hat{X}^\dagger(t_1) \hat{X}(t) >< \hat{D}^\dagger(t') \hat{D}(t_2) \hat{D}^\dagger(t_1) \hat{D}(t) > \tag{29}$$

$$J_{i\to f}^{(intR)} = |U_i|^2 |U_f|^2 \int_{-\infty}^{+\infty} d(t-t') \int_{-\infty}^{t} dt_1 \int_{t'}^{+\infty} dt_2 \\ 2Re\Big[ e^{-i\nu_i (t_1 - t_2)} e^{i\nu_f (t-t')} \\ < \hat{X}^\dagger(t') \hat{X}(t_2) \hat{X}(t) \hat{X}^\dagger(t_1) >< \hat{D}^\dagger(t') \hat{D}(t_2) \hat{D}(t) \hat{D}^\dagger(t_1) > \Big] \tag{30}$$

The physical meaning of these contributions can be understood by noticing the order in which the operators $\hat{D}$ and $\hat{D}^\dagger$ appear in the integrals. In $J_{i\to f}^{(nR)}$ the system starts and ends in the molecular ground state, while in $J_{i\to f}^{(iR)}$ it starts and ends with the molecule in the excited state. These terms describe "normal Raman" and "inverse Raman" scattering processes, respectively.[44] When the molecule is in the ground state only $J_{i\to f}^{(nR)}$ is different from zero while if it is in the excited state only $J_{i\to f}^{(iR)}$ survives. Both contributions exist for molecules with finite probabilities to be in both states, a situation encountered in strongly biased molecular junctions.

In the latter case the third term, $J_{i\to f}^{(intR)}$, results from interference between these two scattering channels. The existence such interference term is a single molecule property, expected to vanish in a thermal steady state ensemble. Note that standard Raman scattering theory starting with a given distribution of molecular electronic states yields the isolated molecule limits of Eqs. (28) and (29) with weights given by this distribution. The interference (30) is not obtained in such treatment that disregards the dynamics associated in our case with the electron and energy exchange with metal leads. (See Section 4 for further discussion).

Eqs. (28)-(30) can be cast in terms of more familiar quantities that will facilitate their estimates below, by noting that the total scattering flux can be written in terms of $J_{i\to f}$ in the form $J_{tot} = \int d\nu_i \int d\nu_f \rho_R(\nu_i) \rho_R(\nu_f) J_{i\to f}$. The integrand, $\rho_R(\nu_i) \rho_R(\nu_f) J_{i\to f}$, is the differential flux, per unit incoming and unit outgoing frequency. Using for $U_i$ and $U_f$ the semiclassical forms $\mathcal{E}\mu_{12}$ where $\mathcal{E}$ is the electric field associated with the radiation field we find that the different contributions to the integrand are given by equations similar to (28)-(30) except that the term $|U_i|^2|U_f|^2$ is replaced by $\left|\mathcal{E}(\nu_i)\mathcal{E}(\nu_f)\right|^2 \Gamma_R(\nu_i)\Gamma_R(\nu_f)/4\pi^2$ where $\Gamma_R(\nu) = 2\pi|\mu_{12}|^2 \rho_R(\nu)$ is the width associated with the $2 \to 1$ radiative relaxation. The electric field terms characterize the important aspect of the local electromagnetic field including possible enhancement effects, but they are not the focus of our present discussion. In what follows we define

$$\rho_R(\nu_i)\rho_R(\nu_f)J_{i\to f} = \left|\mathcal{E}(\nu_i)\mathcal{E}(\nu_f)\right|^2 \bar{J}_{\nu_i\to\nu_f} \tag{31}$$

$$\bar{J}_{\nu_i\to\nu_f} = \frac{\Gamma_R(\nu_i)\Gamma_R(\nu_f)J_{i\to f}}{4\pi^2\,|U_i|^2|U_f|^2} \tag{32}$$

so that the different contributions to $\bar{J}_{\nu_i\to\nu_f}$ are given by equations similar to (28)-(30) where $|U_i|^2|U_f|^2$ is replaced by $\Gamma_R(\nu_i)\Gamma_R(\nu_f)/4\pi^2$.

The Raman flux terms (28)-(30) contain correlation functions in the molecular polarization operators $\hat{D}$ ($\hat{D}^\dagger$) and in the phonon shift operators $\hat{X}$ ($\hat{X}^\dagger$). The latter are dynamical generalizations of standard Franck-Condon factors, and we refer to them as generalized Franck-Condon (GFC) functions. In what follows we outline the ways by which these correlation functions are evaluated.

*Evaluation of the GFC functions.* To simplify the evaluation of these vibrational correlation functions we assume that they can be associated with a thermal distribution characterized by a temperature that reflects the non-equilibrium state of the junction. A way to estimate this vibrational temperature $T_\upsilon$ is described below. In the evaluation itself we disregard in the Hamiltonian (12) the coupling $\hat{V}^{(\upsilon-b)}$ between the molecular vibration and the thermal bath, and expand the correlation function in the basis of free vibrations. For example the GFC function that appears in the normal Raman flux, Eq.(28), is (other GFC factors are calculated similarly)

$$\begin{aligned}&<\hat{X}(t_2)\hat{X}^\dagger(t')\hat{X}(t)\hat{X}^\dagger(t_1)>= \\ &\quad\sum_{\ell_0,\ell,m,n} P_0(\ell_0)<\ell_0|\hat{X}^\dagger|\ell><\ell|\hat{X}|m><m|\hat{X}^\dagger|n><n|\hat{X}|\ell_0> \\ &\quad\times\exp\left(i\omega_\upsilon\left[(\ell_0-\ell)t_2+(\ell-m)t'+(m-n)t+(n-\ell_0)t_1\right]\right)\end{aligned} \tag{33}$$

where $P_0(\ell_0)$ is the equilibrium probability of populating the vibrational level $\ell_0$

$$P_0(\ell_0) = \left[1 - e^{-\omega_\upsilon/T_\upsilon}\right] e^{-\omega_\upsilon \ell_0/T_\upsilon} \tag{34}$$

and where the matrix elements of the shift operator in the free oscillator basis are given by[45]

$$\begin{aligned} &< m \,|\, \hat{X} \,|\, n >= (-1)^{(m-n)\theta(m-n)} \lambda^{|m-n|} \sqrt{\frac{\min(m,n)!}{\max(m,n)!}} e^{-\lambda^2/2} L^{|m-n|}_{\min(m,n)}(\lambda^2) \\ &< m \,|\, \hat{X}^\dagger \,|\, n >= \left[< n \,|\, \hat{X} \,|\, m >\right]^* =< n \,|\, \hat{X} \,|\, m > \end{aligned} \tag{35}$$

where $\theta(x)$ is the step-function and $L_n^\gamma$ are Laguerre polynomials.

*The vibrational temperature.* Next consider the vibrational temperature $T_\upsilon$. In the unbiased junction and in the absence of optical driving the molecular vibrations are assumed to be in equilibrium with the thermal bath ($T_\upsilon = T$). The calculation outlined below assumes that a thermal distribution with some finite temperature persists also in the biased and irradiated junction and relies on two simplifications: We assume that the incident radiation field is weak and does not affect this temperature, and we disregard the effect of coupling to electron-hole pair excitations, $\hat{V}^{(e-h)}$ of Eq. (6). At the steady state driven by a bias potential the rate of junction heating by the electron flux, $J_e$, is equal to the rate of junction cooling by the phonon flux $J_\upsilon$ (due to coupling to thermal bath, $\hat{V}^{(\upsilon-b)}$). These fluxes are given by(for detailed discussion see Ref. [36])

$$J_e = \sum_{K=L,R} \sum_{m=1,2} \int_{-\infty}^{+\infty} \frac{dE}{2\pi} E\left[\Sigma_m^{(K)<}(E) G_m^>(E) - \Sigma_m^{(K)>}(E) G_m^<(E)\right] \tag{36}$$

$$J_\upsilon = -\int_0^\infty \frac{d\omega}{2\pi} \omega\left[\Pi_\upsilon^<(\omega) D_\upsilon^>(\omega) - \Pi_\upsilon^>(\omega) D_\upsilon^<(\omega)\right] \tag{37}$$

and at steady state

$$J_e + J_\upsilon = 0 \tag{38}$$

Here $\Sigma_m^{(K)>,<}(E)$ is the greater (lesser) SE of the electronic orbital $m$ due to coupling to lead $K$

$$\Sigma_m^{(K)<}(E) \quad = i\Gamma_m^{(K)}(E) f_K(E) \tag{39}$$

$$\Sigma_m^{(K)>}(E) \quad = -i\Gamma_m^{(K)}(E)[1 - f_K(E)] \tag{40}$$

where $f_K(E) = [\exp(\beta(E - \mu_K)) + 1]^{-1}$ are Fermi distribution functions and $\Gamma_m^{(K)}(E) = 2\pi \sum_{k \in K} |V_{mk}^{(et)}|^2 \, \delta(E - \varepsilon_k)$. (In writing Eqs. (36) and (39)-(40) we have assumed that the spacing between levels 1 and 2, i.e. the HOMO-LUMO gap, is large relative to their widths, so that non-diagonal elements, $\Sigma_{i,j}^{(K)>,<}(E)$ with $i \neq j$, can be ignored). $\Pi_{\upsilon}^{>}(\omega)$ ($\Pi_{\upsilon}^{<}(\omega)$) are the greater (lesser) SEs of the molecular vibration $\upsilon$ due to coupling to the thermal bath

$$\Pi_{\upsilon}^{<}(\omega) \quad = -i\Omega_{\upsilon}(\omega) f_{\upsilon}(\omega) \tag{41}$$

$$\Pi_{\upsilon}^{>}(\omega) \quad = -i\Omega_{\upsilon}(\omega) f_{\upsilon}(-\omega) \tag{42}$$

with $f_{\upsilon}(\omega) = N_{BE}(\omega)$ for $\omega > 0$ and $1 + N_{BE}(|\omega|)$ for $\omega < 0$, where $N_{BE}(\omega) = [\exp(\beta\omega) - 1]^{-1}$ is the Bose-Einstein distribution in the thermal bath, and $\Omega_{\upsilon}(\omega) = 2\pi \sum_{\beta} |U_{\beta}|^2 \delta(\omega - \omega_{\beta})$. In the wide band approximation invoked in our calculation, where $\Gamma_m^{(K)}(E)$ and $\Omega_{\upsilon}(\omega)$ are assumed constants, they represent respectively the width of molecular level $m$ ($=1,2$) due to coupling to lead $K$ ($=L,R$) and the damping rate of the molecular vibration due to coupling to the thermal bath. To evaluate the currents (36) and (37) we also need the electron and phonon greater and lesser GFs. These are written using yet another simplification including phonon contribution into electronic GF at the Born approximation level and disregarding the electronic contribution to the vibrational GFs. We assume that these corrections to the electronic and vibrational GFs are not very important for the temperature estimate, since they do not change essentially the amount of energy transfered from the electron flux to the vibrational subsystem.

Under these simplifications the zero order single electron GFs, i.e. electronic GFs that enter electronic SE due to phonons within the Born approximation are given by[46]

$$G_m^{<}(E) \quad = i \frac{\Gamma_m^{(L)}(E) f_L(E) + \Gamma_m^{(R)}(E) f_R(E)}{(E - \varepsilon_m)^2 + (\Gamma_m(E)/2)^2} \tag{43}$$

$$G_m^{>}(E) \quad = -i \frac{\Gamma_m^{(L)}(E)[1 - f_L(E)] + \Gamma_m^{(R)}(E)[1 - f_R(E)]}{(E - \varepsilon_m)^2 + (\Gamma_m(E)/2)^2} \tag{44}$$

with $\Gamma_m(E) = \Gamma_m^{(L)}(E) + \Gamma_m^{(R)}(E)$. The molecular vibration GFs (in the quasi-particle approximation) take the form

$$D_{\upsilon}^{<}(\omega) = -2\pi i \left[ N_{\upsilon} \delta(\omega - \omega_{\upsilon}) + (1 + N_{\upsilon}) \delta(\omega + \omega_{\upsilon}) \right] \tag{45}$$

$$D_{\upsilon}^{>}(\omega) = -2\pi i \left[ N_{\upsilon} \delta(\omega + \omega_{\upsilon}) + (1 + N_{\upsilon}) \delta(\omega - \omega_{\upsilon}) \right] \tag{46}$$

Substituting (39)-(46) into (36) and (37) and using (38) we get

$$N_\upsilon = \frac{\Omega_\upsilon N_{BE}(\omega_\upsilon) + I_-}{\Omega_\upsilon + (I_+ - I_-)} \tag{47}$$

$$I_\pm \equiv \sum_{m=1,2} |V_m^{(e-\upsilon)}|^2 \int_{-\infty}^{+\infty} \frac{dE}{2\pi} G_m^<(E) G_m^>(E \pm \omega_v) \tag{48}$$

Expression (47) is used to get the vibrational temperature $T_\upsilon$ under the assumption[47]

$$N_\upsilon = [\exp \omega_\upsilon / T_\upsilon - 1]^{-1} \tag{49}$$

*The polarization correlation functions.* Next consider the molecular polarization correlation functions that enter the expressions for the Raman fluxes (28)-(30). The evaluation of these correlation functions is complicated by the fact that the polarization operators $\hat{D}$ and $\hat{D}^\dagger$, Eq. (10a), are not the true Bose operators. An approximate evaluation proceeds by making two simplifications. First, radiative level broadening is disregarded, i.e. the radiation field is taken just to provide source and drain for photons via the coupling factors $|U_i|^2$ and $|U_f|^2$ in (28)-(30), while the corresponding damping is disregarded relative to the other sources of level broadening in this model. In contrast, the coupling to electron-hole excitations in the leads, represented by $\hat{V}^{(e-h)}$ term in the Hamiltonian (12), is an important ingredient of the physics of molecules near metal surface. It competes for electrons in the excited molecular states with the Raman process and can therefore influence the Raman signal significantly. To account for damping due to this energy relaxation process we make a second approximation employing an ansatz similar to that used in the literature previously[48, 49]

$$\hat{D}(t) \equiv e^{i\hat{\hat{H}}t} \hat{D} e^{-i\hat{\hat{H}}t} \approx e^{i\hat{\hat{H}}^{(et)}t} \hat{D} e^{-i\hat{\hat{H}}^{(et)}t} e^{-\Gamma^{(e-h)}t} \tag{50}$$

where

$$\hat{\hat{H}}^{(et)} \equiv \hat{\hat{H}}_0 + \hat{\hat{V}}^{(et)} \tag{51}$$

is the part of the Hamiltonian (12) including electron transfer only, and where[48, 49]

$$\Gamma^{(e-h)}(E) = 2\pi \sum_{k_1 \neq k_2} |V_{k_1 k_2}^{(e-h)}|^2 f_{k_1}[1 - f_{k_2}] \delta(E - (\varepsilon_{k_2} - \varepsilon_{k_1})) \tag{52}$$

is the molecular polarization damping rate due to coupling to electron-hole excitations in the leads. In the wide ($e-h$ excitations) band approximation this damping function becomes the constant $\Gamma^{(e-h)} \equiv \Gamma^{(e-h)}(\varepsilon_2 - \varepsilon_1)$. This way of taking damping due to electron-hole excitations in the leads into account would be exact (within the wide-band approximation) if there is no electron transfer between leads and molecule (in this case

$\hat{D}$ behaves as a true Bose operator).[50]

After the ansatz is employed the remaining time dependence of $\hat{D}$ is determined by the time evolution $e^{i\hat{\bar{H}}^{(et)}t}\hat{D}e^{-i\hat{\bar{H}}^{(et)}t}$. This time dependence can be made explicit by making some more simplifications. First, electron-phonon coupling is disregarded at this stage of the calculation (it is already accounted for by the correlation functions of the $\hat{X}$ operators in (28)-(30)). Second, as before, mixing of the molecular electronic levels 1 and 2 by their mutual interaction with the leads is disregarded. Under these approximations the polarization correlation functions that appear in Eqs. (28)-(30) are obtained in the form[51]

$$\begin{aligned}
<\hat{D}(t_2)\hat{D}^\dagger(t')\hat{D}(t)\hat{D}^\dagger(t_1)> &\approx e^{-\Gamma^{(e-h)}(t_1-t+t_2-t')/2} \\
&\times[G_1^<(t_1-t_2)G_1^>(t'-t)-G_1^<(t'-t_2)G_1^<(t_1-t)] \\
&\times[G_2^>(t_2-t_1)G_2^<(t-t')-G_2^>(t_2-t')G_2^>(t-t_1)]
\end{aligned} \tag{53}$$

$$\begin{aligned}
<\hat{D}^\dagger(t')\hat{D}(t_2)\hat{D}^\dagger(t_1)\hat{D}(t)> &\approx e^{-\Gamma^{(e-h)}(t_1-t+t_2-t')/2} \\
&\times[G_1^<(t_1-t_2)G_1^>(t'-t)-G_1^>(t'-t_2)G_1^>(t_1-t)] \\
&\times[G_2^>(t_2-t_1)G_2^<(t-t')-G_2^<(t_2-t')G_2^<(t-t_1)]
\end{aligned} \tag{54}$$

$$\begin{aligned}
<\hat{D}^\dagger(t')\hat{D}(t_2)\hat{D}(t)\hat{D}^\dagger(t_1)> &\approx e^{-\Gamma^{(e-h)}(t_1-t+t_2-t')/2} \\
&\times[G_1^<(t_1-t_2)G_1^>(t'-t)-G_1^>(t'-t_2)G_1^<(t_1-t)] \\
&\times[G_2^>(t_2-t_1)G_2^<(t-t')-G_2^<(t_2-t')G_2^>(t-t_1)]
\end{aligned} \tag{55}$$

where $G_m^{>,<}(t)$ $(m=1,2)$ are the Fourier transform of the single electron GFs given by Eqs. (43) and (44).

Finally, utilizing (50) and substituting (33) and (53)-(55) into (28)-(32), we get (after transforming the single electron GFs to the energy domain) the final expressions for Raman scattering fluxes used in our calculations.[52]

$$\bar{J}^{(nR)}_{\nu_i\to\nu_f} = \frac{\Gamma_R(\nu_i)\Gamma_R(\nu_f)}{2\pi}\sum_{\ell_0,m} P_0(\ell_0)$$

$$\Bigg\{\delta(\nu_i+\omega_\upsilon\ell_0-\nu_f-\omega_\upsilon m)\Bigg|\sum_n\int\frac{dE^{(1)}}{2\pi}\int\frac{dE^{(2)}}{2\pi}G_1^<(E^{(1)})G_2^>(E^{(2)})$$

$$\times\frac{<m|\hat{X}^\dagger|n><n|\hat{X}|\ell_0>}{\nu_i+E^{(1)}+\omega_\upsilon\ell_0-E^{(2)}-\omega_\upsilon n+i\Gamma^{(e-h)}/2}\Bigg|^2$$

$$+\int\frac{dE_1^{(1)}}{2\pi}\int\frac{dE_2^{(1)}}{2\pi}\delta(\nu_i+\omega_\upsilon\ell_0-\nu_f-\omega_\upsilon m+E_1^{(1)}-E_2^{(1)})G_1^<(E_1^{(1)})G_1^>(E_2^{(1)})$$

$$\times\Bigg|\sum_n\int\frac{dE^{(2)}}{2\pi}G_2^>(E^{(2)})\frac{<m|\hat{X}^\dagger|n><n|\hat{X}|\ell_0>}{\nu_i+E_1^{(1)}+\omega_\upsilon\ell_0-E^{(2)}-\omega_\upsilon n+i\Gamma^{(e-h)}/2}\Bigg|^2$$

$$+\int\frac{dE_1^{(2)}}{2\pi}\int\frac{dE_2^{(2)}}{2\pi}\delta(\nu_i+\omega_\upsilon\ell_0-\nu_f-\omega_\upsilon m-E_1^{(2)}+E_2^{(2)})G_2^>(E_1^{(2)})G_2^<(E_2^{(2)})$$

$$\times\Bigg|\sum_n\int\frac{dE^{(1)}}{2\pi}G_1^<(E^{(1)})\frac{<m|\hat{X}^\dagger|n><n|\hat{X}|\ell_0>}{\nu_i+E^{(1)}+\omega_\upsilon\ell_0-E_1^{(2)}-\omega_\upsilon n+i\Gamma^{(e-h)}/2}\Bigg|^2$$

$$+\int\frac{dE_1^{(1)}}{2\pi}\int\frac{dE_2^{(1)}}{2\pi}\int\frac{dE_1^{(2)}}{2\pi}\int\frac{dE_2^{(2)}}{2\pi}G_1^<(E_1^{(1)})G_1^>(E_2^{(1)})G_2^>(E_1^{(2)})G_2^<(E_2^{(2)})$$

$$\times\delta(\nu_i+\omega_\upsilon\ell_0-\nu_f-\omega_\upsilon m+E_1^{(1)}-E_2^{(1)}-E_1^{(2)}+E_2^{(2)}) \qquad (56)$$

$$\times\Bigg|\sum_n\frac{<m|\hat{X}^\dagger|n><n|\hat{X}|\ell_0>}{\nu_i+E_1^{(1)}+\omega_\upsilon\ell_0-E_1^{(2)}-\omega_\upsilon n+i\Gamma^{(e-h)}/2}\Bigg|^2\Bigg\}$$

$$\bar{J}^{(iR)}_{\nu_i\to\nu_f} = \frac{\Gamma_R\left(\nu_i\right)\Gamma_R\left(\nu_f\right)}{2\pi}\sum_{\ell_0,m} P_0(\ell_0)$$

$$\Bigg\{\delta(\nu_i+\omega_\upsilon\ell_0-\nu_f-\omega_\upsilon m)\left|\sum_n\int\frac{dE^{(1)}}{2\pi}\int\frac{dE^{(2)}}{2\pi}G_1^>(E^{(1)})G_2^<(E^{(2)})\right.$$

$$\left.\times\frac{<\ell_0\,|\,\hat{X}\,|\,n><n\,|\,\hat{X}^\dagger\,|\,m>}{\nu_i+E^{(1)}+\omega_\upsilon n-E^{(2)}-\omega_\upsilon m+i\Gamma^{(e-h)}/2}\right|^2$$

$$+\int\frac{dE_1^{(1)}}{2\pi}\int\frac{dE_2^{(1)}}{2\pi}\delta(\nu_i+\omega_\upsilon\ell_0-\nu_f-\omega_\upsilon m-E_1^{(1)}+E_2^{(1)})G_1^>(E_1^{(1)})G_1^<(E_2^{(1)})$$

$$\times\left|\sum_n\int\frac{dE^{(2)}}{2\pi}G_2^<(E^{(2)})\frac{<\ell_0\,|\,\hat{X}\,|\,n><n\,|\,\hat{X}^\dagger\,|\,m>}{\nu_i+E_2^{(1)}+\omega_\upsilon n-E^{(2)}-\omega_\upsilon m+i\Gamma^{(e-h)}/2}\right|^2$$

$$+\int\frac{dE_1^{(2)}}{2\pi}\int\frac{dE_2^{(2)}}{2\pi}\delta(\nu_i+\omega_\upsilon\ell_0-\nu_f-\omega_\upsilon m+E_1^{(2)}-E_2^{(2)})G_2^<(E_1^{(2)})G_2^>(E_2^{(2)})$$

$$\times\left|\sum_n\int\frac{dE^{(1)}}{2\pi}G_1^>(E^{(1)})\frac{<\ell_0\,|\,\hat{X}\,|\,n><n\,|\,\hat{X}^\dagger\,|\,m>}{\nu_i+E^{(1)}+\omega_\upsilon n-E_2^{(2)}-\omega_\upsilon m+i\Gamma^{(e-h)}/2}\right|^2$$

$$+\int\frac{dE_1^{(1)}}{2\pi}\int\frac{dE_2^{(1)}}{2\pi}\int\frac{dE_1^{(2)}}{2\pi}\int\frac{dE_2^{(2)}}{2\pi}G_1^>(E_1^{(1)})G_1^<(E_2^{(1)})G_2^<(E_1^{(2)})G_2^>(E_2^{(2)})$$

$$\times\delta(\nu_i+\omega_\upsilon\ell_0-\nu_f-\omega_\upsilon m-E_1^{(1)}+E_2^{(1)}+E_1^{(2)}-E_2^{(2)}) \qquad (57)$$

$$\times\left|\sum_n\frac{<\ell_0\,|\,\hat{X}\,|\,n><n\,|\,\hat{X}^\dagger\,|\,m>}{\nu_i+E_2^{(1)}+\omega_\upsilon n-E_2^{(2)}-\omega_\upsilon m+i\Gamma^{(e-h)}/2}\right|^2\Bigg\}$$

$$
\begin{aligned}
\bar{J}^{(intR)}_{\nu_i\to\nu_f} = \frac{\Gamma_R(\nu_i)\Gamma_R(\nu_f)}{2\pi}\sum_{\ell_0,m} P_0(\ell_0)\times 2Re\Big\{ & -\delta(\nu_i+\omega_\upsilon\ell_0-\nu_f-\omega_\upsilon m) \\
& \times\sum_{\ell}\int\frac{dE_1^{(1)}}{2\pi}\int\frac{dE_1^{(2)}}{2\pi}G_1^{>}(E_1^{(1)})G_2^{<}(E_1^{(2)}) \\
& \times\frac{<\ell_0|\hat{X}|\ell><\ell|\hat{X}^{\dagger}|m>}{\nu_i+E_1^{(1)}+\omega_\upsilon\ell-E_1^{(2)}-\omega_\upsilon m+i\Gamma^{(e-h)}/2} \\
& \times\sum_{n}\int\frac{dE_2^{(1)}}{2\pi}\int\frac{dE_2^{(2)}}{2\pi}G_1^{<}(E_2^{(1)})G_2^{>}(E_2^{(2)}) \\
& \times\frac{<m|\hat{X}^{\dagger}|n><n|\hat{X}|\ell_0>}{\nu_i+E_2^{(1)}+\omega_\upsilon\ell_0-E_2^{(2)}-\omega_\upsilon n+i\Gamma^{(e-h)}/2} \\
& +\int\frac{dE_1^{(1)}}{2\pi}\int\frac{dE_2^{(1)}}{2\pi}\delta(\nu_i+\omega_\upsilon\ell_0-\nu_f-\omega_\upsilon m-E_1^{(1)}+E_2^{(1)})G_1^{>}(E_1^{(1)})G_1^{<}(E_2^{(1)}) \\
& \times\sum_{\ell}\int\frac{dE_1^{(2)}}{2\pi}G_2^{<}(E_1^{(2)})\frac{<\ell_0|\hat{X}|\ell><\ell|\hat{X}^{\dagger}|m>}{\nu_i+E_2^{(1)}+\omega_\upsilon\ell-E_1^{(2)}-\omega_\upsilon m+i\Gamma^{(e-h)}/2} \\
& \times\sum_{n}\int\frac{dE_2^{(2)}}{2\pi}G_2^{>}(E^{(2)})\frac{<m|\hat{X}^{\dagger}|n><n|\hat{X}|\ell_0>}{\nu_i+E_2^{(1)}+\omega_\upsilon\ell_0-E_2^{(2)}-\omega_\upsilon n+i\Gamma^{(e-h)}/2} \\
& +\int\frac{dE_1^{(2)}}{2\pi}\int\frac{dE_2^{(2)}}{2\pi}\delta(\nu_i+\omega_\upsilon\ell_0-\nu_f-\omega_\upsilon m+E_1^{(2)}-E_2^{(2)})G_2^{<}(E_1^{(2)})G_2^{>}(E_2^{(2)}) \\
& \times\sum_{\ell}\int\frac{dE_1^{(1)}}{2\pi}G_1^{>}(E_1^{(1)})\frac{<\ell_0|\hat{X}|\ell><\ell|\hat{X}^{\dagger}|m>}{\nu_i+E_1^{(1)}+\omega_\upsilon\ell-E_2^{(2)}-\omega_\upsilon m+i\Gamma^{(e-h)}/2} \\
& \times\sum_{n}\int\frac{dE_2^{(1)}}{2\pi}G_1^{<}(E_2^{(1)})\frac{<m|\hat{X}^{\dagger}|n><n|\hat{X}|\ell_0>}{\nu_i+E_2^{(1)}+\omega_\upsilon\ell_0-E_2^{(2)}-\omega_\upsilon n+i\Gamma^{(e-h)}/2} \\
& -\int\frac{dE_1^{(1)}}{2\pi}\int\frac{dE_2^{(1)}}{2\pi}\int\frac{dE_1^{(2)}}{2\pi}\int\frac{dE_2^{(2)}}{2\pi}G_1^{>}(E_1^{(1)})G_1^{<}(E_2^{(1)})G_2^{<}(E_1^{(2)})G_2^{>}(E_2^{(2)}) \\
& \times\delta(\nu_i+\omega_\upsilon\ell_0-\nu_f-\omega_\upsilon m-E_1^{(1)}+E_2^{(1)}+E_1^{(2)}-E_2^{(2)}) \\
& \times\sum_{\ell}\frac{<\ell_0|\hat{X}|\ell><\ell|\hat{X}^{\dagger}|m>}{\nu_i+E^{(1)2}+\omega_\upsilon\ell-E_2^{(2)}-\omega_\upsilon m+i\Gamma^{(e-h)}/2} \\
& \times\sum_{n}\frac{<m|\hat{X}^{\dagger}|n><n|\hat{X}|\ell_0>}{\nu_i+E_2^{(1)}+\omega_\upsilon\ell_0-E_2^{(2)}-\omega_\upsilon n+i\Gamma^{(e-h)}/2}\Big\}
\end{aligned}
\qquad (58)
$$

To end this discussion we consider the limit of an isolated molecule, where $\Gamma_m^{(K)}\to 0$ ($K=L,R$) and $\Gamma^{(e-h)}\to 0$. In this limit Green functions $G_m^{>,<}(E)$ ($m=1,2$) become $G_m^{<}(E)=in_m\delta(E-\varepsilon_m)$ and $G_m^{>}(E)=-i[1-n_m]\delta(E-\varepsilon_m)$. Using these in Eq. (58) yields zero. Eq. (56) leads to

$$\bar{J}^{(nR)}_{\nu_i\to\nu_f} \to \frac{\Gamma_R(\nu_i)\Gamma_R(\nu_f)}{2\pi}\sum_{\ell_0,m} P_0\delta(\nu_i+\omega_\upsilon\ell_0-\nu_f-\omega_\upsilon m)$$
$$\times\left|\frac{<m|\hat{X}^\dagger|n><n|\hat{X}|\ell_0>}{\nu_i+\varepsilon_1+\omega_\upsilon\ell_0-\varepsilon_2-\omega_\upsilon n+i\delta}\right|^2\times F(n_1,n_2) \tag{59}$$

where $n_m$ ($m=1,2$) are the average level populations and all four terms in Eq. (56) combine to give

$$\begin{aligned}F(n_1,n_2) &= n_1^2[1-n_2]^2+n_1[1-n_1][1-n_2]^2+n_1^2n_2[1-n_2]+n_1[1-n_1]n_2[1-n_2]\\ &= n_1[1-n_2]\end{aligned}. \tag{60}$$

Apart from the factor $n_1[1-n_2]$ this is the standard expression for normal Raman scattering. The additional factor is the probability to find the molecule in the state that allows normal Raman scattering, i.e. occupied ground state and empty excited state. Obviously, for an isolated molecule at room temperature this factor is 1. Similarly, Eq. (57) yields a term proportional to $n_2[1-n_1]$ that is zero for an isolated molecule.

The above results were obtained for model M. In the general case charge transfer can occur with phonon excitation or de-excitation, the total radiative interaction (7) may be written in the form

$$\hat{V}^{(e-p)} = \sum_{\alpha\in i,\{f\}}\left(\hat{a}_\alpha^\dagger\hat{O}_\alpha+\hat{O}_\alpha^\dagger\hat{a}_\alpha\right) \tag{61}$$

$$\hat{O}_\alpha = U_\alpha^{(e-p)}\hat{D}+\sum_{k\in\{L,R\}}\sum_{m=1,2}\left(V_{mk,\alpha}^{(e-p)}\hat{D}_{mk}+V_{km,\alpha}^{(e-p)}\hat{D}_{km}\right) \tag{62}$$

where the operators $\hat{D}, \hat{D}_{mk}, \hat{D}_{km}$ were defined in Eqs. (10). After the small polaron transformation this becomes

$$\begin{aligned}\hat{O}_\alpha &\to U_\alpha^{(e-p)}\hat{D}\hat{X}+\sum_{k\in\{L,R\}}\left(V_{1k,\alpha}^{(e-p)}\hat{D}_{1k}\hat{X}_1^\dagger+V_{k2,\alpha}^{(e-p)}\hat{D}_{k2}\hat{X}_2+V_{k1,\alpha}^{(e-p)}\hat{D}_{k1}\hat{X}_1+V_{2k,\alpha}^{(e-p)}\hat{D}_{2k}\hat{X}_2^\dagger\right)\\ &\equiv \hat{O}_\alpha^{(M)}+\hat{O}_\alpha^{(1)}+\hat{O}_\alpha^{(2)}+\hat{O}_\alpha^{(3)}+\hat{O}_\alpha^{(4)}\end{aligned} \tag{63}$$

where $\hat{X}_m=\exp\left(i\lambda_m\hat{P}_v\right)$ and $\hat{X}=\hat{X}_1^\dagger\hat{X}_2$. The formal evaluation proceeds as before, with the analog of Eq. (26) for the state to state photon flux taking the form

$$J_{i\to f} = \int_{-\infty}^{\infty} d(t-t')\int_c d\tau_1\int_c d\tau_2\Pi_f^>(t-t')\Pi_i(\tau_1,\tau_2)\left\langle T_c\hat{O}_f^\dagger(t')\hat{O}_f(t)\hat{O}_i^\dagger(\tau_1)\hat{O}_i(\tau_2)\right\rangle \tag{64}$$

With the operators $\hat{O}$ given by (63) we are facing the need to evaluate $5^4$ = 625

integrals. In the low bias regime where electronic state 1 is occupied while 2 is unoccupied, the terms $\hat{O}_\alpha^{(3)} + \hat{O}_\alpha^{(4)}$ that are associated with $V_{k1,\alpha}^{(e-p)} \hat{D}_{k1} \hat{X}_1 + V_{2k,\alpha}^{(e-p)} \hat{D}_{2k} \hat{X}_2^\dagger$ may be disregarded reducing the number of integrals to $3^4 = 81$. Eqs. (56)-(58) are obtained when the charge transfer contributions could be disregarded in (63), i.e. when $\hat{O}_\alpha = \hat{O}_\alpha^{(M)}$ in Eq. (64). We expect that this is the case when the incident radiation is close to resonance with the molecular transition.

Away from this resonance the charge transfer components of $\hat{O}$ may be important. In particular, $\hat{O}_\alpha^{(1)} = \sum_{k\in\{L,R\}} \sum_{m=1,2} V_{1k,\alpha}^{(e-p)} \hat{D}_{1k} \hat{X}_1^\dagger$ describes transition between the molecular HOMO and the metal while $\hat{O}_\alpha^{(2)} = \sum_{k\in\{L,R\}} \sum_{m=1,2} V_{k2,\alpha}^{(e-p)} \hat{D}_{k2} \hat{X}_2$ represents transtions between the metal and the molecular LUMO. If we take $\hat{O}_\alpha = \hat{O}_\alpha^{(1)}$ or $\hat{O}_\alpha = \hat{O}_\alpha^{(2)}$ in Eq. (64) and limit ourself to the 'normal' Raman component that dominates the signal at low bias, we will get contributions to the Raman scattering analogous to the molecule-to-metal and metal-to-molecule transitions, respectively, of Ref. [9]. In general, however, all terms in Eq. (63) should be taken into account together in Eq. (64), but the evaluation of this general expression is not feasible within the present formalism. Instead we will study the particular case where the metal-to-molecule charge-transfer transition is assumed to dominate the Raman signal, i.e. where $\hat{O}_\alpha$ is replaced by $\hat{O}_\alpha^{(2)}$ in Eq. (64). (The equivalent case where the molecule-to-metal transition dominates can be evaluated in the same way). Furthermore, we focus on the normal Raman process. This will make it possible for us to make contact with the theory of Ref. [9] and to compare the processes associated with the molecular excitation represented by $\hat{O}_\alpha^{(M)}$ and with the charge transfer transition.

Using

$$\hat{O}_\alpha = \hat{O}_\alpha^{(2)} = \sum_{k\in\{L,R\}} \sum_{m=1,2} V_{k2,\alpha}^{(e-p)} \hat{D}_{k2} \hat{X}_2 \tag{65}$$

in Eq. (64) and repeating the calculation that leads to Eq. (28), yields the following form for normal Raman component of this contribution

$$\begin{aligned} J_{i\to f}^{(nR)} = \sum_{k,k',k_1,k_2\in\{L,R\}} & V_{k_2 2,i}^{(e-p)} V_{2k',f}^{(e-p)} V_{k2,f}^{(e-p)} V_{2k_1,i}^{(e-p)} \int_{-\infty}^{+\infty} d(t-t') \int_{-\infty}^{t} dt_1 \int_{-\infty}^{t'} dt_2 \, e^{-i\nu_i (t_1 - t_2)} e^{i\nu_f (t-t')} \\ & \times \left\langle \hat{X}_2(t_2) \hat{X}_2^\dagger(t') \hat{X}_2(t) \hat{X}_2^\dagger(t_1) \right\rangle \left\langle \hat{c}_{k_2}^\dagger(t_2) \hat{d}_2(t_2) \hat{d}_2^\dagger(t') \hat{c}_{k'}(t') \hat{c}_k^\dagger(t) \hat{d}_2(t) \hat{d}_2^\dagger(t_1) \hat{c}_{k_1}(t_1) \right\rangle \end{aligned} \tag{66}$$

The average of the product of nuclear displacement operators is done as before. The electronic average is first approximated as a product $\left\langle \hat{c}^{\dagger}_{k_2}(t_2)\hat{c}_{k'}(t')\hat{c}^{\dagger}_{k}(t)\hat{c}_{k_1}(t_1)\right\rangle \left\langle \hat{d}_2(t_2)\hat{d}^{\dagger}_2(t')\hat{d}_2(t)\hat{d}^{\dagger}_2(t_1)\right\rangle$. This simplification is based on our intuitive expectation that the main contributions to the sums over the $k$ indices will come from energy regime far (by ~ $\hbar\nu_i$ or $\hbar\nu_f$) from $\varepsilon_2$, so that the corresponding single electron states are not appreciably mixed by the molecule-metal charge-transfer interaction. Using Wick's theorem then leads to

$$\left\langle \hat{c}^{\dagger}_{k_2}(t_2)\hat{c}_{k'}(t')\hat{c}^{\dagger}_{k}(t)\hat{c}_{k_1}(t_1)\right\rangle = \delta_{k,k'}\delta_{k_1,k_2}g^{<}_{k_1}(t_1-t_2)\,g^{>}_{k}(t'-t) - \delta_{k,k_1}\delta_{k',k_2}g^{<}_{k}(t_1-t)\,g^{<}_{k'}(t'-t_2) \tag{67}$$

where $g^{<}_{k}(t) = if_K(\varepsilon_k)e^{-i\varepsilon_k t}$ and $g^{>}_{k}(t) = -i\left[1-f_K(\varepsilon_k)\right]e^{-i\varepsilon_k t}$ for $k \in K$ are the free electron GFs in the metal, and

$$\left\langle \hat{d}_2(t_2)\hat{d}^{\dagger}_2(t')\hat{d}_2(t)\hat{d}^{\dagger}_2(t_1)\right\rangle = G^{>}_2(t_2-t_1)\,G^{<}_2(t-t') - G^{>}_2(t_2-t')\,G^{>}_2(t-t_1) \tag{68}$$

Using these in (66) leads to the analog of Eq. (56) for the metal-to-molecule normal Raman charge transfer process

$$\bar{J}^{(nR)}_{\nu_i\to\nu_f} = \frac{2}{\pi}\sum_{\ell_0,m} P_0(\ell_0)$$

$$\left\{ \delta(\nu_i + \omega_\upsilon \ell_0 - \nu_f - \omega_\upsilon m)\left| \sum_n \int \frac{dE^{(1)}}{2\pi}\int\frac{dE^{(2)}}{2\pi}\sum_{K=L,R}\bar{S}^{(K),<}_{2,if}(E^{(1)})G_2^{>}(E^{(2)}) \right.\right.$$

$$\left.\times\frac{<m|\hat{X}_2^\dagger|n><n|\hat{X}_2|\ell_0>}{\nu_i + E^{(1)}+\omega_\upsilon\ell_0 - E^{(2)} - \omega_\upsilon n + i\Gamma^{(e-h)}/2}\right|^2$$

$$+\int\frac{dE_1^{(1)}}{2\pi}\int\frac{dE_2^{(1)}}{2\pi}\delta(\nu_i+\omega_\upsilon\ell_0-\nu_f-\omega_\upsilon m + E_1^{(1)} - E_2^{(1)})$$

$$\times\left(\sum_{K=L,R}\bar{S}^{(K),<}_{2,ii}(E_1^{(1)})\right)\left(\sum_{K=L,R}\bar{S}^{(K),>}_{2,ff}(E_2^{(1)})\right)$$

$$\times\left|\sum_n\int\frac{dE^{(2)}}{2\pi}G_2^{>}(E^{(2)})\frac{<m|\hat{X}_2^\dagger|n><n|\hat{X}_2|\ell_0>}{\nu_i+E_1^{(1)}+\omega_\upsilon\ell_0-E^{(2)}-\omega_\upsilon n+i\Gamma^{(e-h)}/2}\right|^2$$

$$+\int\frac{dE_1^{(2)}}{2\pi}\int\frac{dE_2^{(2)}}{2\pi}\delta(\nu_i+\omega_\upsilon\ell_0-\nu_f-\omega_\upsilon m - E_1^{(2)} + E_2^{(2)})G_2^{>}(E_1^{(2)})G_2^{<}(E_2^{(2)})$$

$$\times\left|\sum_n\int\frac{dE^{(1)}}{2\pi}\sum_{K=L,R}\bar{S}^{(K),<}_{2,if}(E^{(1)})\frac{<m|\hat{X}_2^\dagger|n><n|\hat{X}_2|\ell_0>}{\nu_i+E^{(1)}+\omega_\upsilon\ell_0-E_1^{(2)}-\omega_\upsilon n+i\Gamma^{(e-h)}/2}\right|^2$$

$$+\int\frac{dE_1^{(1)}}{2\pi}\int\frac{dE_2^{(1)}}{2\pi}\int\frac{dE_1^{(2)}}{2\pi}\int\frac{dE_2^{(2)}}{2\pi}\delta(\nu_i+\omega_\upsilon\ell_0-\nu_f-\omega_\upsilon m+E_1^{(1)}-E_2^{(1)}-E_1^{(2)}+E_2^{(2)})$$

$$\times\left(\sum_K\bar{S}^{(K),<}_{2,ii}(E_1^{(1)})\right)\left(\sum_K\bar{S}^{(K),>}_{2,ff}(E_2^{(1)})\right)G_2^{>}(E_1^{(2)})G_2^{<}(E_2^{(2)})$$

$$\left.\times\left|\sum_n\frac{<m|\hat{X}_2^\dagger|n><n|\hat{X}_2|\ell_0>}{\nu_i+E_1^{(1)}+\omega_\upsilon\ell_0-E_1^{(2)}-\omega_\upsilon n+i\Gamma^{(e-h)}/2}\right|^2\right\} \tag{69}$$

where the functions $\bar{S}^{(K)}_{2,\alpha\alpha'}(E)$ are related to the self energy-like functions associated with the radiative metal-to-molecule charge transfer. On the Keldysh contour the latter functions are

$$S^{(K)}_{2,\alpha\alpha'}(\tau,\tau') = \sum_{k\in K} V^{(e-p)}_{2k,\alpha}\, g_k(\tau,\tau')V^{(e-p)}_{k2,\alpha'} \tag{70}$$

or disregarding the dependence on the indices *α* and *α'* as discussed in the paragraph following Eq. (10)

$$S_2^{(K)<}(E) = \sum_{k\in K} V^{(e-p)}_{2k} g_k^{<}(E) V^{(e-p)}_{k2} = i\Upsilon_2^{(K)}(E) f_K(E) \tag{71}$$

$$S_2^{(K)>}(E) = \sum_{k\in K} V^{(e-p)}_{2k} g_k^{>}(E) V^{(e-p)}_{k2} = -i\Upsilon_2^{(K)}(E)\left[1 - f_K(E)\right] \tag{72}$$

Here we have used $g_k^{<}(E) = 2\pi i f_K(E)\delta(E-\varepsilon_k)$ and $g_k^{>}(E) = -2\pi i[1-f_K(E)]\delta(E-\varepsilon_k)$ and have defined

$$\Upsilon_2^{(K)}(E) = 2\pi \sum_{k\in K} \left|V_{2k}^{(e-p)}\right|^2 \delta(E-\varepsilon_k) = 2\pi \left( \left|V_{2k}^{(e-p)}\right|^2 \rho_K \right)_E . \tag{73}$$

Where $\rho_K$ is the density of free electron states in the lead K. The corresponding function $\bar{S}_2^{(K)}(E)$ is defined so as to implement a transformation similar to (32)

$$\bar{S}_2^{(K)<}(E) = iC_2^{(K)}(E) f_K(E); \qquad \bar{S}_2^{(K)>}(E) = -iC_2^{(K)}(E)[1-f_K(E)] \tag{74}$$

where[53]

$$C_2^{(K)}(E) = \Upsilon_2^{(K)}(E)\rho_R \tag{75}$$

Note that Eq. (69) has a form similar to (55), except that the shift operator $\hat{X}$ is replaced by $\hat{X}_2$ and the Green function $G_1$ is replaced by $\sum_{K=L,R} \bar{S}_2^{(K)}$. Also note that in this calculation we disregard the difference between $\rho_R(\nu_i)$ and $\rho_R(\nu_f)$

Below, we refer to results based on Eqs. (64), (65) and (69) as Model CT, keeping in mind that this is just a representative contribution of the charge transfer mechanism to Raman scattering by molecules adsorbed on metals. To make it possible to compare the pure molecular and the charge transfer contributions, Eqs. (56) and (69), respectively, we need to use some reasonable estimate of the couplings involved. To this end we adopt a model due to Persson, in which the optical charge-transfer interaction, $\hat{V}_{CT}^{(e-p)}$, is taken to be (for the molecular level 2)

$$\hat{V}_{CT}^{(e-p)} = e\delta E \hat{d}_2^\dagger \hat{d}_2 \tag{76}$$

where $\delta$ is the metal-molecule distance, $E$ is the electric field associated with the radiation field (essentially the analog of the operator $\hat{a}_\alpha + \hat{a}_\alpha^\dagger$ in Eq. (7c)) and $e$ is the electron charge. The interaction (76) represents modulation of the molecular energy (relative to the metal Fermi energy) by the radiation field by an amount equal to the work needed to move an electron between metal and molecule under this field. To bring it to the form (7c) we write $\hat{d}_2$ as a linear combination of diagonal state operators, $\hat{d}_2 = \sum_j K_j \hat{c}_j$ ($\{j\}$ are the exact single electron states associated with the molecular level 2, the metal and the interaction (4) between them, and $\hat{c}_j$ are the corresponding single particle operators). Furthermore, anticipating that the most contributing metal states are near the Fermi energy that is assumed to be far from level 2, we make the approximation

$$\sum_j K_j \hat{\tilde{c}}_j = \sum_k \frac{V^{(et)}}{\Delta E} \hat{c}_k \tag{77}$$

where $V^{(et)}$ is the coupling between level 2 and states near the metal Fermi energy and $\Delta E = \varepsilon_2 - E_F$ is the separation between these energies. Eq. (76) may then be written in the form

$$\hat{V}_{CT}^{(e-p)} = e\delta E \frac{V^{(et)}}{\Delta E} \sum_k \hat{d}_2^\dagger \hat{c}_k + h.c. \tag{78}$$

Consider now the parameter $C/(2\pi) \equiv \left\{ \left| \left\langle 2 | \hat{V}_{CT}^{(e-p)} | k \right\rangle \right|^2 \rho\rho_R \right\}_{E_F}$, where ρ is the density of metal states near the Fermi energy and $\rho_R$ is the density of radiation field modes. Note that this is the function (75) evaluated at $E = E_F$ for the lead under consideration. This parameter measures the coupling strength for the radiative transition between molecular state 2 and the continuum of metal states at the Fermi energy. Using $(e\delta E)^2 \rho_R \sim \Gamma_R$, where $\Gamma_R$ is the radiative emission rate for a molecule that couples to the radiation field with a transition dipole $e\delta$, and $\left(V^{(et)}\right)^2 \rho \sim \Gamma$, the electron transfer rate between molecule to metal, leads to

$$C/(2\pi) \equiv \left\{ \left| \left\langle 2 | \hat{V}_{C}^{(e-p)} | k \right\rangle \right|^2 \rho\rho_R \right\}_{E_F} = \frac{\Gamma_R}{\Gamma} \left( \frac{\Gamma}{\Delta E} \right)^2 \tag{79}$$

Below we use a generic value $\Gamma_R / \hbar = 10^9 \, s^{-1}$ to provide an order of magnitude estimate for the parameter C, that will depend via $\Delta E$ on an imposed bias.

## 4. Results and discussion

In the calculations described below we used Eqs. (56)-(58) for the normal and inverse Raman signals and their interference. For the same model, Eq. (12), we also calculate the current-voltage characteristic according to the procedure described in Refs. [27, 28]. These calculations are done using a symmetric voltage division factor for the applied bias, i.e.

$$\mu_K = E_F + \eta_K eV \quad (K = L, R) \qquad \eta_L = 1 + \eta_R = 0.5 \tag{80}$$

Numerical integrations are done using an energy grid that spans the range from $-2$ to 2

eV with step $10^{-4}$ eV. To make these calculations feasible we have assumed that these integrals are dominated by regions of their variables where $E_1^{(1)} = E_2^{(1)}$ and $E_1^{(2)} = E_2^{(2)}$, e.g. we have replaced delta functions such as $\delta(\nu_i + \omega_\upsilon \ell_0 - \nu_f - \omega_\upsilon m + E_1^{(2)} - E_2^{(2)})$ by $\delta(\nu_i + \omega_\upsilon \ell_0 - \nu_f - \omega_\upsilon m)$. As seen from the structure of the Green functions terms, this approximation is valid only for resonance excitation, so that our off resonance calculations below should be regarded of qualitative value only.

In most of these calculations we consider a symmetric junction with a "standard" set of parameters $E_F = 0, \varepsilon_1 = -1$ eV, $\varepsilon_2 = 1$ eV, $\omega_\upsilon = 0.1$ eV, $\Omega_\upsilon = 0.005$ eV and $T = 100$ K. Other choices of these parameters are indicated specifically below. For the molecule-leads coupling $\Gamma_m$ and $V_m^{(e-\upsilon)}$ $(m = 1,2)$ we note that different choices reflect different weak and strong coupling scenarios considered in earlier works. In particular, the mechanism described by model CT may be important when the molecule is chemically bonded to the metal $(\Gamma_m \simeq 0.1 - 1\text{eV})$, while model M is expected to be dominant when the molecule is separated from the metal (a retracted STM tip or a metal substrate covered by a thin insulating layer), where $\Gamma_m$ is considerably smaller. Also note that the vibronic coupling parameters $V_m^{(e-\upsilon)}$ reflect the reorganization energies associated with molecular charging (electron transfer between metal and electronic orbital *m*), while their difference, $\left|V_1^{(e-\upsilon)} - V_2^{(e-\upsilon)}\right|$ corresponds to nuclear reorganization associated with molecular excitation and is responsible for the Raman signal in the isolated molecule (and in model M). Particular choices for these parameters are indicated below.

Two types of Raman scattering signals are described below. The energy resolved scattering from $\nu_i$ to $\nu_f$ is given by the flux $\bar{J}_{\nu_i \to \nu_f}$. To account for finite energy resolution this is obtained from Eqs. (56)-(58) by replacing $\delta$-functions by

$$\delta(x) \to \frac{1}{\pi} \frac{\delta}{x^2 + \delta^2} \tag{81}$$

with $\delta$ chosen to be $0.001$ eV. The integral over this energy resolved signal is the total Raman scattering intensity,

$$\bar{J}_{\nu_i \to \{\nu_f\}} = \int d\nu_f \bar{J}_{\nu_i \to \nu_f} \tag{82}$$

Finally, some of the figures below display the different contributions from the normal, inverse, and interference terms, Eqs. (56)-(58), as well as Stokes ( $\nu_f = \nu_i - \omega_\upsilon$) and anti-Stokes ( $\nu_f = \nu_i + \omega_\upsilon$) signals.

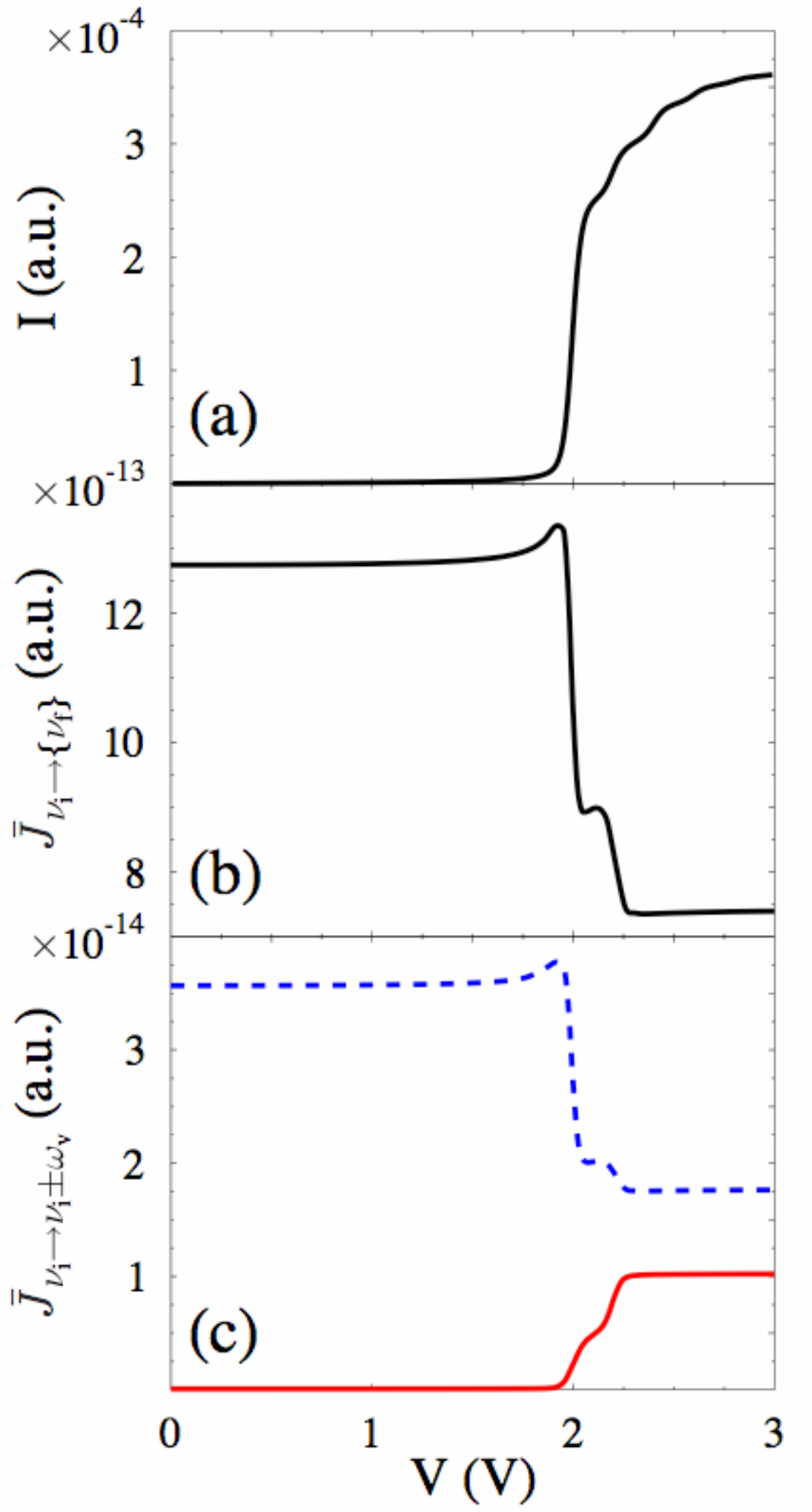

Figure 1: (Color online) (a) The current $I$ , (b) the integrated Raman signal $\bar{J}_{\nu_i \to \{\nu_f\}}$, and (c) the Stokes $\bar{J}_{\nu_i \to \nu_i - \omega_\upsilon}$ (dashed line, blue) and anti-Stokes $\bar{J}_{\nu_i \to \nu_i + \omega_\upsilon}$ (solid line, red) intensities displayed as functions of the applied bias $V$ for resonance incident light, $\nu_i = \varepsilon_2 - \varepsilon_1$. In addition to the standard parameters (see text), the following electronic and vibronic coupling parameters are used here: $\Gamma_m = 0.01\,\text{eV}$ ($m = 1,2$), $\Gamma^{(e-h)} = 0.01\,\text{eV}$, $V_1^{(e-\upsilon)} = 0.1$ eV and $V_2^{(e-\upsilon)} = 0.05$ eV. Here and below "a.u." stands for atomic units.

We start with model M. Figure 1 shows the total Raman scattering, $\bar{J}_{\nu_i \to \{\nu_f\}}$, as well as the Stokes, $\bar{J}_{\nu_i \to \nu_i - \omega_\upsilon}$, and anti-Stokes, $\bar{J}_{\nu_i \to \nu_i + \omega_\upsilon}$, components of the energy resolved signal displayed against the source-drain voltage $V$ . Also shown is the current $I$ across the junction. At the voltage threshold for conduction, $V = 2\,\text{V}$, where the molecular levels enter the window between the chemical potentials on the leads, current through the junction (Fig. 1a) increases due to resonance tunneling through these levels. The modulation of this current by the molecular vibration is manifested by the set of steps in the current right above this threshold. The total Raman signal (Fig. 1b) drops by almost half at the same threshold. This can be understood from the following argument. Disregarding for now the interference contribution (it will be discussed below) the total Raman signal is approximately a sum of the normal and inverse contributions. For an isolated molecule (or a molecule weakly coupled to the leads) the former is proportional to $n_1(1-n_2)$ ($n_m$, $m=1,2$ are the average populations of the molecular levels), while the latter is proportional to $n_2(1-n_1)$. Below threshold the HOMO is populated, $n_1=1$, while LUMO is empty $n_2=0$, so the contribution to total signal comes only from normal Raman process with weighting factor $n_1(1-n_2)=1$. Well above threshold $n_1 \approx n_2 \approx 0.5$, and the total signal consists of normal and inverse Raman contributions each entering with weighting factor $n_1(1-n_2) \approx n_2(1-n_1) \approx 1/4$. Assuming that the two contributions are equal, the total weighting factor is $1/2$, which explains the drop of the Raman signal by such a factor above the threshold. As is seen in Fig. 1c the Stokes intensity decreases while the anti-Stokes signal increases beyond threshold. The above argument concerning electronic level populations would by itself imply a decrease of both components beyond threshold, however heating of the molecular vibration by the resonance electronic current contributes towards an overall increase of the anti-Stokes component.

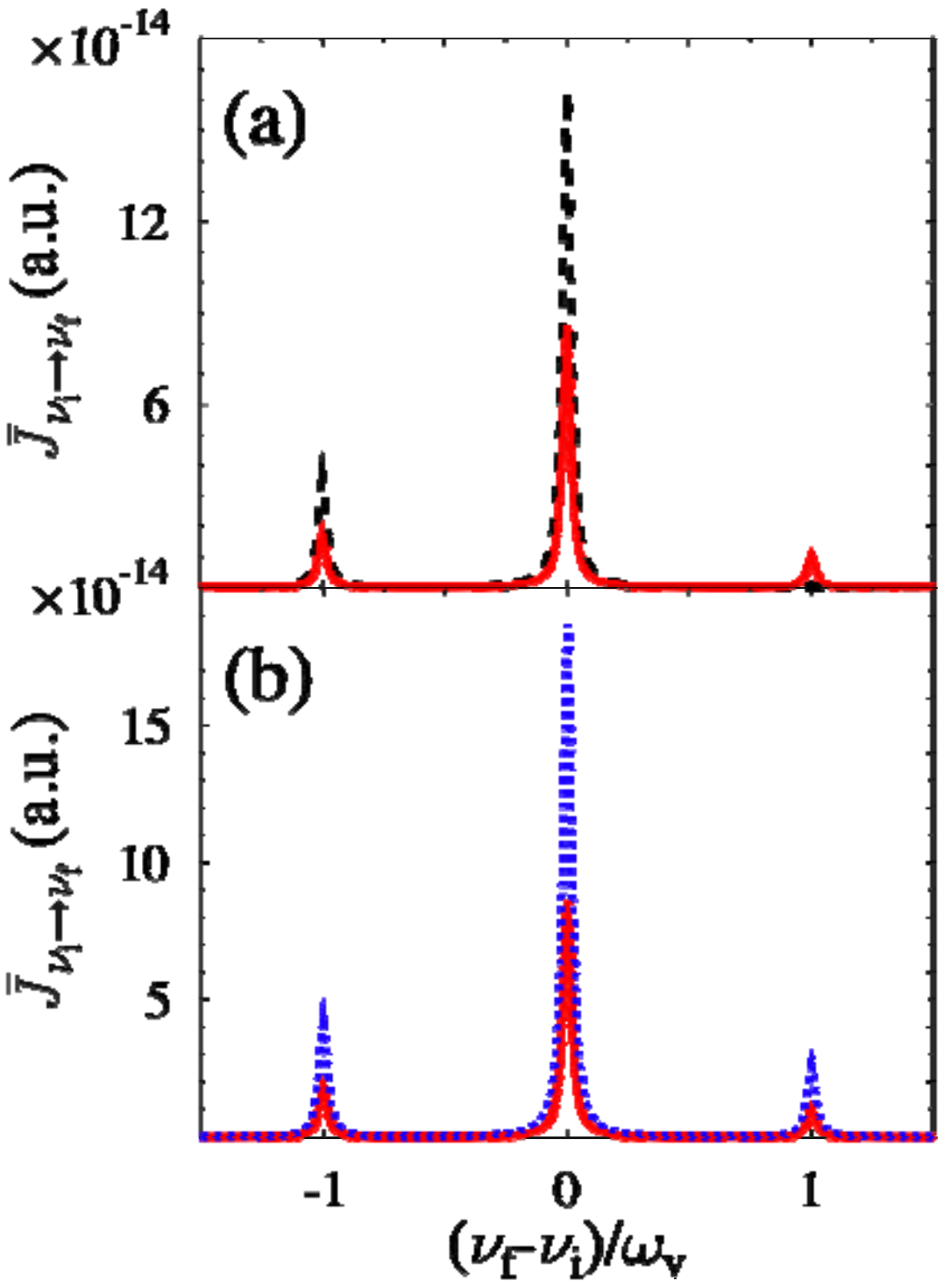

Figure 2: (Color online) Raman signal as function of outgoing frequency $\nu_f$ for fixed incoming frequency at resonance, $\nu_i = \varepsilon_2 - \varepsilon_1 \equiv \Delta\varepsilon$. Fig. 2a shows results at no bias, $V=0$ (dashed line, black) and above threshold $V=2.5$ V (solid line, red). Parameters are the same as in Fig. 1. Fig. 2b compares the cases $\Gamma^{(e-h)}=0.01$ eV (solid line, red) and $\Gamma^{(e-h)}=0.004$ eV (dotted line, blue) for $V=2.5$ V.

Figure 2 shows the energy resolved Raman signal as a function of the outgoing frequency $\nu_f$ for fixed incoming frequency at resonance, $\nu_i = \varepsilon_2 - \varepsilon_1 \equiv \Delta\varepsilon$, for equilibrium, $V = 0$ (dashed line, black), and at bias above the conduction threshold, $V = 2.5$ V (solid line, red). Qualitatively similar spectra are obtained at other bias potentials and in non-resonance situations $|\nu_i - \Delta\varepsilon| \gg \Gamma_m$, however absolute intensities vary strongly. For example, the Raman/Rayleigh (inelastic/elastic) peak ratios at $\nu_i = \Delta\varepsilon$ are 0.219 ($V = 0$), 0.219 ($V = 1.5$ V), 0.204 ($V = 2.5$ V) for the Stokes ($\nu_f = \nu_i - \omega_\upsilon$) signal and $4.242\times10^{-4}$ ($V = 0$), $4.441\times10^{-4}$ ($V1.5$ V), 0.118 ($V = 2.5$ V) for the anti-Stokes ($\nu_f = \nu_i + \omega_\upsilon$) line. For a different choice of vibronic couplings, $V_1^{(e-\upsilon)} = 0.1$ eV and $V_2^{(e-\upsilon)} = 0.08$ eV, the corresponding ratios for resonance excitation are $3.566\times10^{-2}$ ($V = 0$), $3.568\times10^{-2}$ ($V = 1.5$ V), $6.061\times10^{-2}$ ($V = 2.5$ V) for the Stokes and $4.037\times10^{-4}$ ($V = 0$), $4.107\times10^{-4}$ ($V = 1.5$ V), $4.018\times10^{-2}$ ($V = 2.5$ V) for the anti-Stokes intensities. In the off-resonance case both the overall scattering intensity and the relative inelastic signals are considerably smaller. For the case $\nu_i = \Delta\varepsilon/2$, $V_1^{(e-\upsilon)} = 0.1$ eV and $V_2^{(e-\upsilon)} = 0.05$ eV the Raman/Rayleigh peak ratios are $2.318\times10^{-3}$ ($V = 0$), $2.414\times10^{-3}$ ($V = 1.5$ V), 0.108 ($V = 2.5$ V) and $4.001\times10^{-4}$ ($V = 0$), $4.006\times10^{-4}$ ($V = 1.5$ V), 0.094 ($V = 2.5$ V) for the Stokes and anti-Stokes lines, respectively. These numbers depend on the damping parameters $\Gamma_m$ and $\Gamma^{(e-h)}$, as exemplified in Fig. 2b. Again, the anti-Stokes component is significant only above the threshold potential of 2 V due to junction heating.

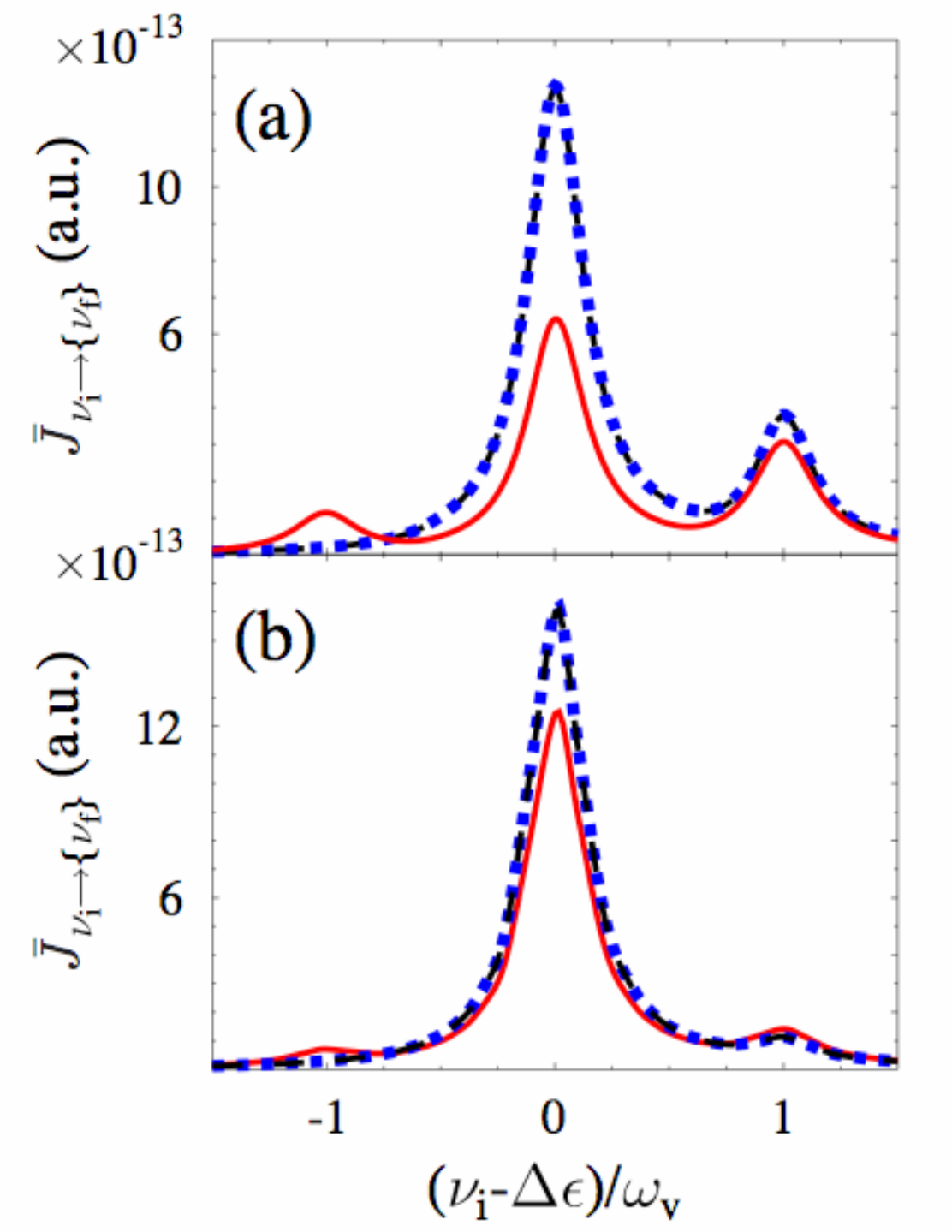

Figure 3: (Color online) The total (integrated) light scattering signal, Eq. (82), displayed as function of the incoming frequency $\nu_i$ for no bias $V = 0$ (dashed line, black) and junction biased below ($V = 1.5$ V, dotted line, blue) and above ($V = 2.5$ V, solid line, red) the conduction threshold. The sub-threshold, $V$=0 and $V$=1.5V, lines lie on top of each other. (a) Parameters as in Fig. 1. (b) Same parameters, except $V_1^{(e-\upsilon)} = 0.1$ eV and $V_2^{(e-\upsilon)} = 0.08$ eV.

The total (integrated) light scattering intensity, essentially the absorption spectrum, is plotted against the incoming frequency $\nu_i$ in Fig. 3 for the cases of unbiased junction $V = 0$ (dashed line, black) as well as junction biased below, $V = 1.5$ V (dotted line, blue) and above, $V = 2.5$ V (solid line, red), the conduction threshold. The absorption lineshape is seen to change considerably above the conduction threshold. Also here we see the increase in the anti-Stokes peak intensities at $\nu_i = \Delta\varepsilon - \omega_\upsilon$ above this threshold.

The existence of the intereference contribution, Eq. (58), to the Raman scattering is an interesting and perhaps surprising result of the present theory that will be elaborated upon elsewhere. Here we limit ourselves to a brief qualitative discussion. This contribution practically vanishes (together with the inverse Raman component) below the conduction threshold (here taken $V = 2$ eV), and clearly depends on the excited state population that forms above this threshold. Having two channels, one that starts and ends with the molecule in the ground state, the other that employs the molecule in the excited state, implies that such interference may take place in analogy to light transmitted through a wall with two slits. Further reflection shows that the situation is more complex. A single molecule described by the first term of the spinless Hamiltonian (2) can be in one of four states: A zero- and a two-electron states may represent molecular cation and anion respectively, and two one-electron states with the electron in the lower or the upper level represent the ground and excited states of the neutral molecule. The energy spacing $\Delta\varepsilon = \varepsilon_2 - \varepsilon_1$ between the single electron levels is taken large relative to $k_B T$ so that in an unbiased junction the molecule is, with probability of essentially 1, in its ground neutral state. When the bias is large enough ($V > \Delta\varepsilon$ in the model considered) all the states are occupied with finite probabilities. Still, within the present model M, Raman processes involve the ground and excited states of the "neutral" molecular species. A detailed examination shows however that interference between the two Raman channels does not arise from the mere fact that both states are populated in the biased junction, but from the correlated dynamical switching between them induced by the electron-hole relaxation process associated with the $V^{(e-h)}$ term (6) in the Hamiltonian (1). This is seen in Figure 4, where the ratio between the interference component and the total Raman Stokes signal is plotted against the electron-hole relaxation rate $\Gamma^{(e-h)}$ and the electron transfer rate $\Gamma_m$. The interference contribution seems to vanish in the limit $\Gamma^{(e-h)} = 0$ as expected. The dependence on $\Gamma_m$ is more complex: A finite $\Gamma_m$ helps to affect population of the excited molecular state (electron in the upper state; hole in the lower state) in the junction. This may explain the initial increase with increasing $\Gamma_m$ in the (negative) interference contribution in Fig. 4b. However as $\Gamma_m$ increases further, the populations of these states are determined by the uncorrelated events of electron exchange between molecule and metal rather then by the correlated switching between the ground and excited molecular states induced by $\Gamma^{(e-h)}$, and the interference contribution diminishes.

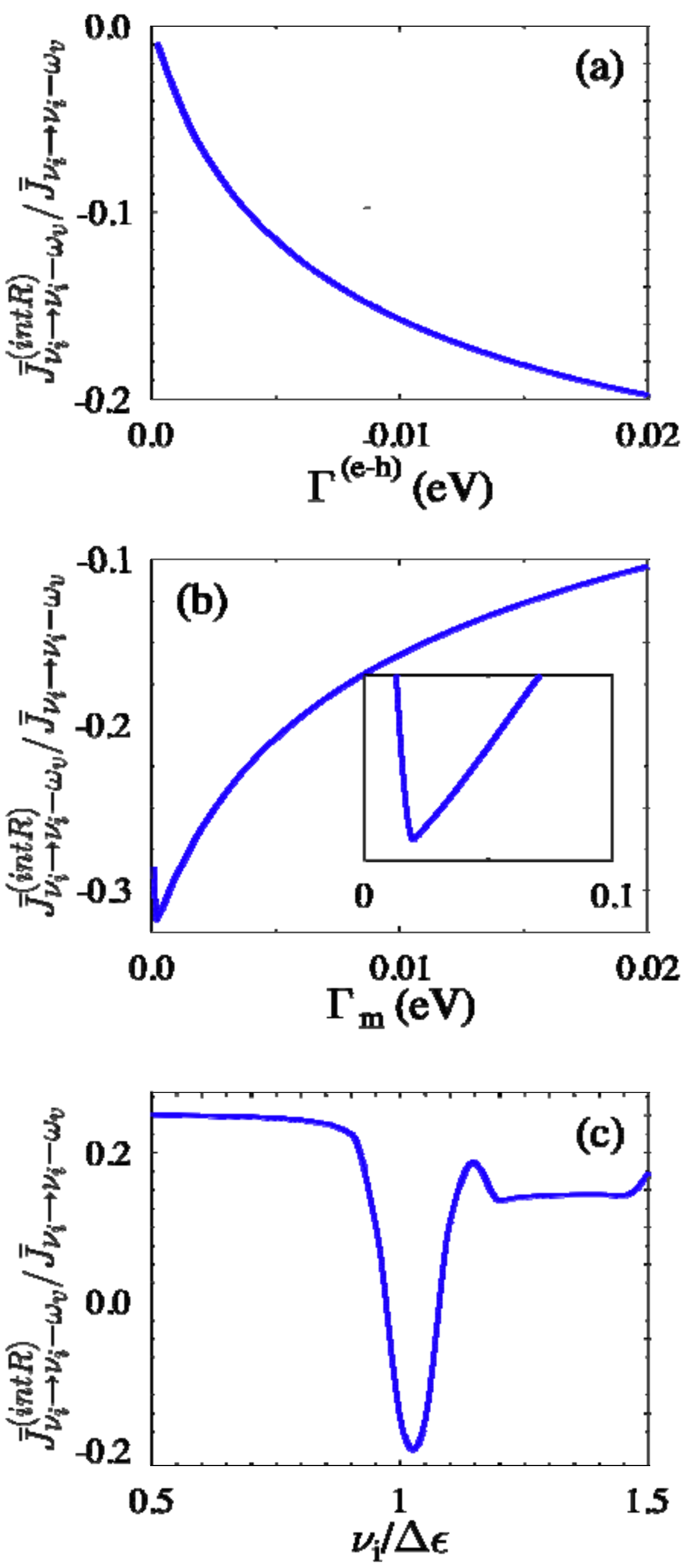

Figure 4: The ratio between the interference component and the total Raman flux at the Stokes frequency $\nu_f = \nu_i - \omega_\upsilon$, $\bar{J}^{(intR)}_{\nu_i \to \nu_i - \omega_\upsilon} / \bar{J}_{\nu_i \to \nu_i - \omega_\upsilon}$, at $V = 2.5$ V as a function of (a) $\Gamma^{(e-h)}$ for $\Gamma_m = 0.01$ eV and $\nu_i = \Delta\varepsilon$; (b) $\Gamma_m$ for $\Gamma^{(e-h)} = 0.01$ eV and $\nu_i = \Delta\varepsilon$ (The inset shows the small $\Gamma_m$ region in detail); (c) $\nu_i$ for $\Gamma_m = \Gamma^{(e-h)} = 0.01$ eV. Parameters are as in Fig. 3b.

Figures 4a and 4b correspond to resonant incident light, $\nu_i = \Delta\varepsilon$. Fig. 4c shows the dependence of the relative interference contribution on the incident light frequency. Interestingly, for our choice of parameters this contribution is negative for resonance excitation and becomes positive for off resonance scattering.

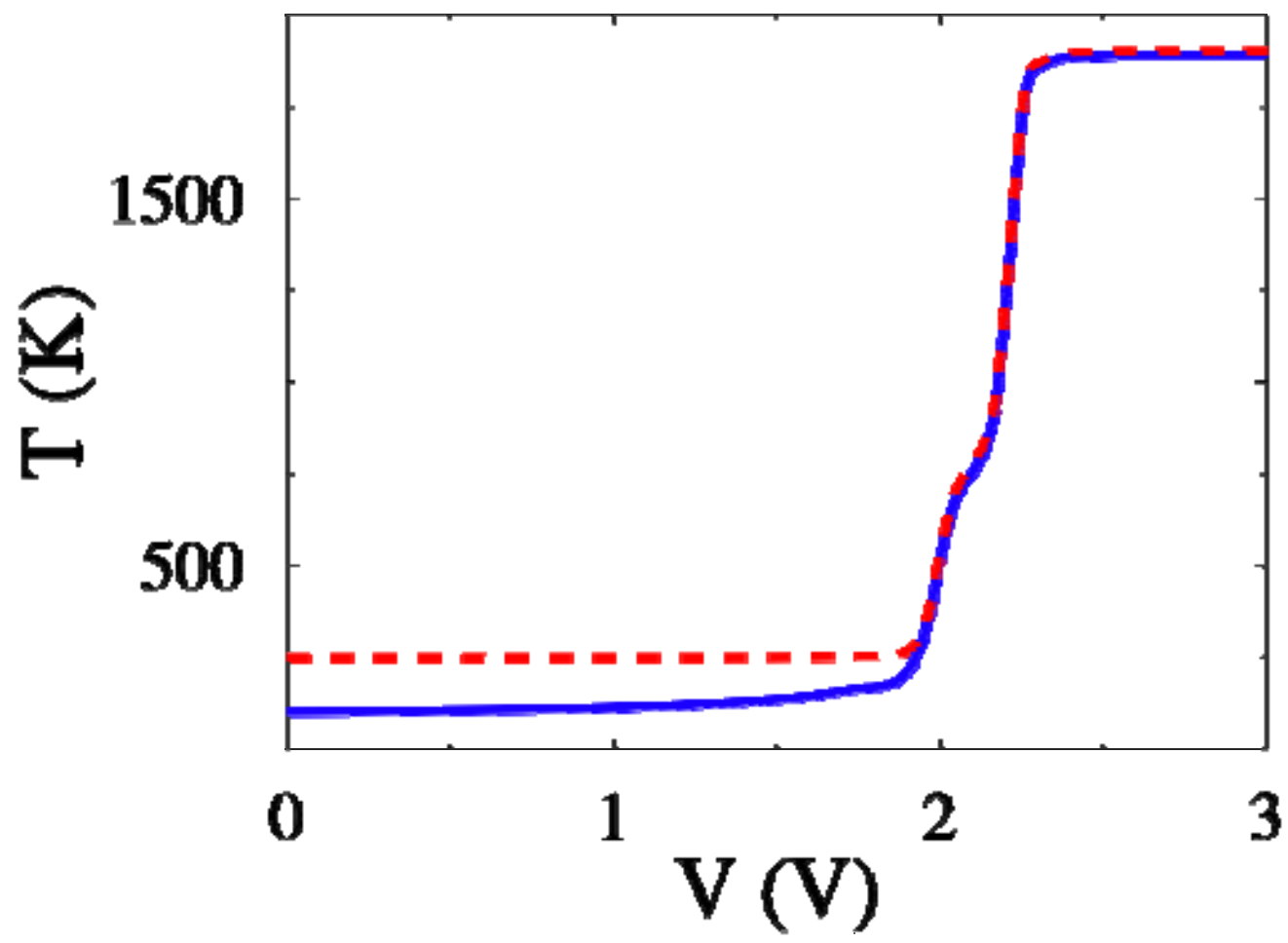

Figure 5: (Color online) Comparison between the vibrational temperature $T_\upsilon$, calculated from Eq. (49) (solid line, blue) and the temperature estimated from the ratio of Stokes and anti-Stokes intensities $T_{S-aS}$, Eq. (83) (dashed line, red) vs. applied bias. Calculation is done for the resonant scattering case $\nu_i = \varepsilon_2 - \varepsilon_1 \equiv \Delta\varepsilon$ . Parameters are as in Fig. 3b

Observation of Raman scattering from molecular junctions is in principle a useful tool for estimating the junction temperature.[50] Indeed, assuming that the ratio between the Stokes and anti-Stokes Raman scattering features reflects the relative populations of the corresponding vibrational levels, and assuming again that these populations are determined by some Boltzmann distribution associated with the junction non-equilibrium temperature, this temperature is obtained from

$$T_{S-aS} = \frac{\omega_\upsilon}{\ln\left(\frac{\bar{J}_{\nu_i\to\nu_i-\omega_\upsilon}}{\bar{J}_{\nu_i\to\nu_i+\omega_\upsilon}} \frac{(\nu_i+\omega_\upsilon)^2}{(\nu_i-\omega_\upsilon)^2}\right)}. \tag{83}$$

The factors $(\nu_i \pm \omega_\upsilon)^2$ correct for the frequency dependence of the outgoing radiation field density of modes. Eq. (83) may be used to characterize heating of the molecular vibration in the biased junction.[54] It should be noted that additional frequency dependent corrections may be caused by resonance structure in the scattering cross-section. Figure 5 compares the temperature of the molecular vibration $T_\upsilon$ obtained from Eq. (49) with the temperature estimated from the Raman Stokes/anti-Stokes ratio according to Eq. (83). While the two temperatures follow the same qualitative behavior, they differ quantitatively from each other at low voltage. This apparent failure is associated with the fact that in the low voltage regime the anti-Stokes signal is negligible, leading to large errors in its estimate.

Next we consider the contribution of metal-molecule charge transfer transitions to the Raman scattering, which in our model originates from the coupling (7c) and exemplified by the particular contribution (69). It was suggested[6,8,9] that this contribution, the so called chemical or first layer effect, may play a significant role in the observed SERS signal, where expected spectral features at $\nu_i \sim E_F - \varepsilon_1$ (molecule to metal charge transfer) or $\nu_i \sim \varepsilon_2 - E_F$ (metal to molecule charge transfer) can be dressed by molecular vibrational transitions.[9] have suggested that the sharp fall in the electronic occupation of metal levels across the Fermi energy leads to a maximum in these contributions to the molecular Raman scattering, a view supported by the observed dependence of SERS signal in electrochemical systems on the incident light frequency at different reference electrode potentials.[55]

To see if and how such effects are manifested by our model we consider first the equilibrium (no bias) case, where (except for details involving surface selection rules) a molecule coupled to two electrodes is equivalent to a molecule adsorbed on a single metal substrate. At room temperature the molecular HOMO, level 1, is occupied and the LUMO, level 2, is empty, and 'normal' Raman scattering prevails. Our result, Eq. (69), for the Raman signal associated with the metal-to-molecule charge transfer transition, should be analogous to the corresponding result of Ref. [9], although the latter was cast in the language of intensity borrowing that, as agrued above,[29] is unnecessary. We note however that in Ref. [9] the molecular levels are assumed to be well represented by those of an isolated molecule, i.e. the electron transfer interaction between molecule and metal, Eq. (4), is disregarded. Instead a phenomenological width parameter is added in the resulting scattering expression (Eq. 26 of Ref. [9]).

To see the similarity as well as the difference between our result and that of Ref. [9] we focus by way of illustration on the first term on right-hand-side of Eq. (69)

$$I \equiv \sum_n < m | \hat{X}_2^\dagger | n >< n | \hat{X}_2 | \ell_0 > \int \frac{dE^{(1)}}{2\pi} \int \frac{dE^{(2)}}{2\pi} \sum_{K=L,R} S_2^{(K)<}(E^{(1)}) G_2^{>}(E^{(2)}) \\ \times \frac{1}{\nu_i + E^{(1)} + \omega_\upsilon \ell_0 - E^{(2)} - \omega_\upsilon n + i\Gamma^{(e-h)}/2} \tag{84}$$

Using (43) and (44), setting $\mu_L = \mu_R = E_F$, and taking $T \to 0$, the integral over $E^{(1)}$ yields

$$I = \sum_n < m | \hat{X}_2^\dagger | n >< n | \hat{X}_2 | \ell_0 > \int \frac{dE^{(2)}}{2\pi} iG_2^>\left(E^{(2)}\right) \\ \times \frac{\Upsilon_2}{2\pi} \ln \frac{E_F - E^{(2)} + \omega_\upsilon(\ell_0 - n) + \nu_i + i\Gamma^{(e-h)}/2}{-D - E^{(2)} + \omega_\upsilon(\ell_0 - n) + \nu_i + i\Gamma^{(e-h)}/2} \tag{85}$$

where $D$ measures the leads half-bandwidth and $\Upsilon_2$ was defined in Eq. (73). When $D$ is larger than all the other relevant energy scales, the logarithm term in the integrand becomes

$$\ln \frac{\sqrt{\left(E_F - E^{(2)} + \omega_\upsilon(\ell_0 - n) + \nu_i\right)^2 + \left(\Gamma^{(e-h)}/2\right)^2}}{D} \tag{86}$$

which indeed gives a peak at $\nu_i = E^{(2)} - E_F - \omega_\upsilon(\ell_0 - n)$. The height of this peak is related to $\Gamma^{(e-h)}/D$, and it becomes more pronounced for smaller $\Gamma^{(e-h)}$. This peak, addressed in Ref. [9], is still subjected to integration over $E^{(2)}$ in (85), and the final result will show a corresponding peak behavior only for sufficiently small $\Gamma_2$. This is shown in Fig. 6a, which depicts the Raman Stokes signal as function of the incident frequency. Obviously, broadening by $\Gamma^{(e\text{-}h)}$ or by the Fermi distribution at higher temperatures will have similar effects on erasing the peak structure. In general, the step structure of the Fermi function can lead to a peak, but only under a relatively strict set of conditions that are sometime questionable. For example, in Fig. 6b the Raman Stokes signal is plotted against $\varepsilon_2$ that reflects dependence on a gate or reference electrode potential. Using the same parameters as in Fig 6a we find a peak structure only when the radiative charge transfer coupling $V^{(e-p)}$ is taken constant, namely, when its dependence on $\varepsilon_2$ acording to Eq. (78) ($\Delta E = \varepsilon_2 - E_F$) is disregarded.

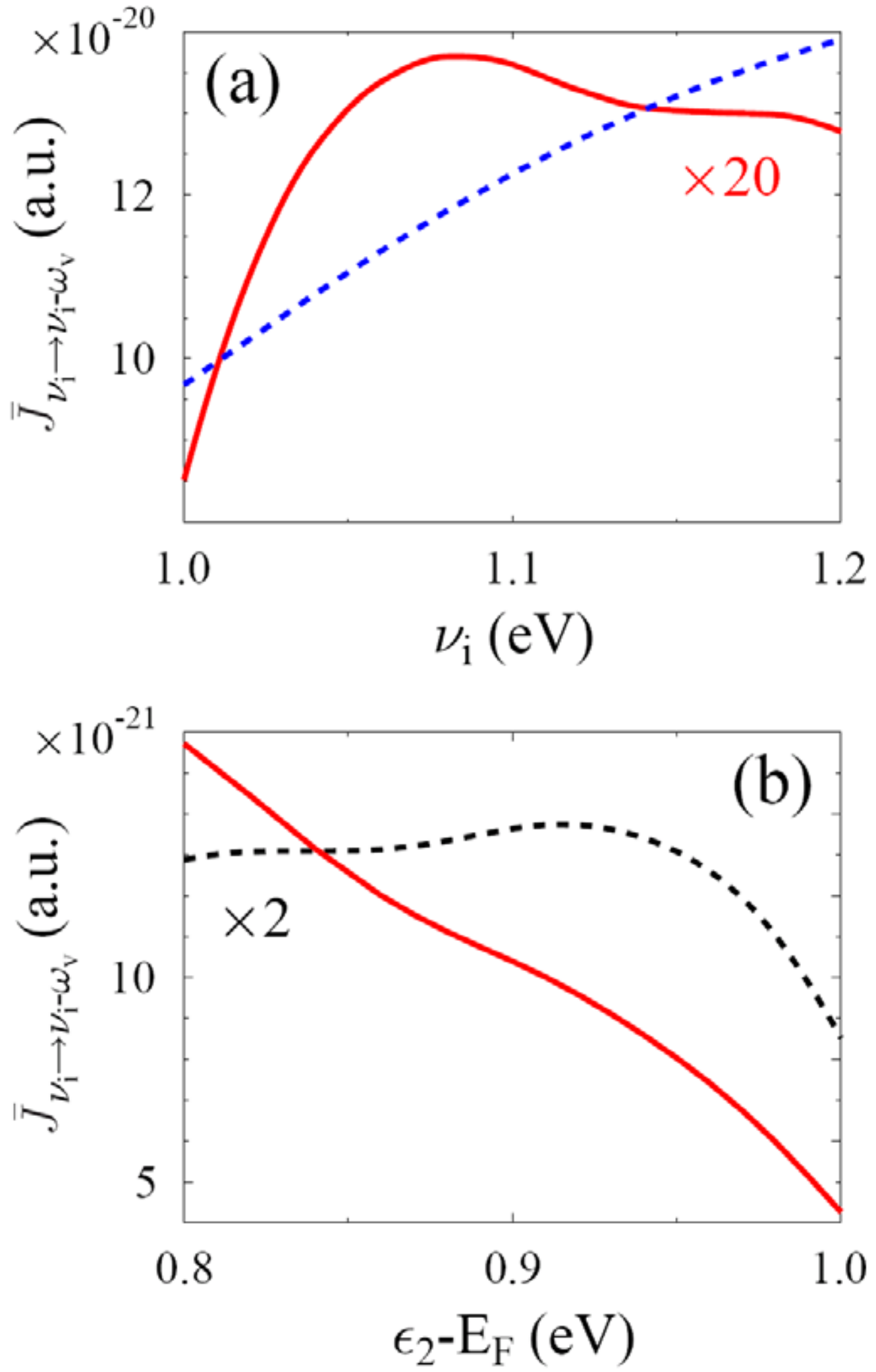

Figure 6: (Color online) (a) The Stokes component of the Raman flux associated with the metal-to-molecule charge transfer transition, plotted against the incident light frequency for the choice of parameters $E_F$ = 0, $\varepsilon_2$ = 1 eV, $\Gamma^{(e-h)} = 0.01\text{eV}$, $\omega_\upsilon$ = 0.1eV, $\Omega_\upsilon$ = 0.005 eV, $V_1^{(e-\upsilon)} = 0.1\text{eV}$, $V_2^{(e-\upsilon)} = 0.08\text{eV}$ ; $T$=100K . The electron transfer coupling corresponds to $\Gamma_2$ = 0.05 eV for the solid line (red) and $\Gamma_2$ = 0.5 eV for the dashed line (blue). (b) The same (Stokes) component evaluated for $\nu_i$ = 1 eV, plotted against $\varepsilon_2$, the position of level 2, which measures the effect of varying reference electrode potential. $\Gamma_2 = 0.05$ eV is used in both lines. The full (red) line results from a calculation for which the radiative charge-transfer coupling $\hat{V}_B^{(e-p)}$ is obtained from Eq. (78). The dashed (black) line is obtained from a similar calculation except that $\hat{V}_B^{(e-p)}$ is taken constant, at the value obtained from Eq. (78) for $\Delta E = \varepsilon_2 - E_F = 1\text{eV}$.

Similar factors affect the light scattering behavior of a molecule connected to two leads in a biased junction, however more structure in the scattering spectrum is expected when the molecule interacts with leads of different Fermi energies. When the bias is small so that $\varepsilon_1 < \mu_L, \mu_R < \varepsilon_2$, the molecular HOMO and LUMO remain occupied and unoccupied, respectively. Structure similar to Fig. 6 is now expected about the energies $\varepsilon_2 - \mu_L$ and $\varepsilon_2 - \mu_R$. Obviously similar argument and similar structures appear near $\mu_L - \varepsilon_1$ and $\varepsilon_1 - \mu_R$, associated with the resonant molecule (HOMO)-to-metal charge transfer transition. Above the conduction threshold, when the LUMO level becomes partly populated the same argument would predict similar features associated with inverse Raman (metal-to-HOMO and LUMO-to-metal) processes.

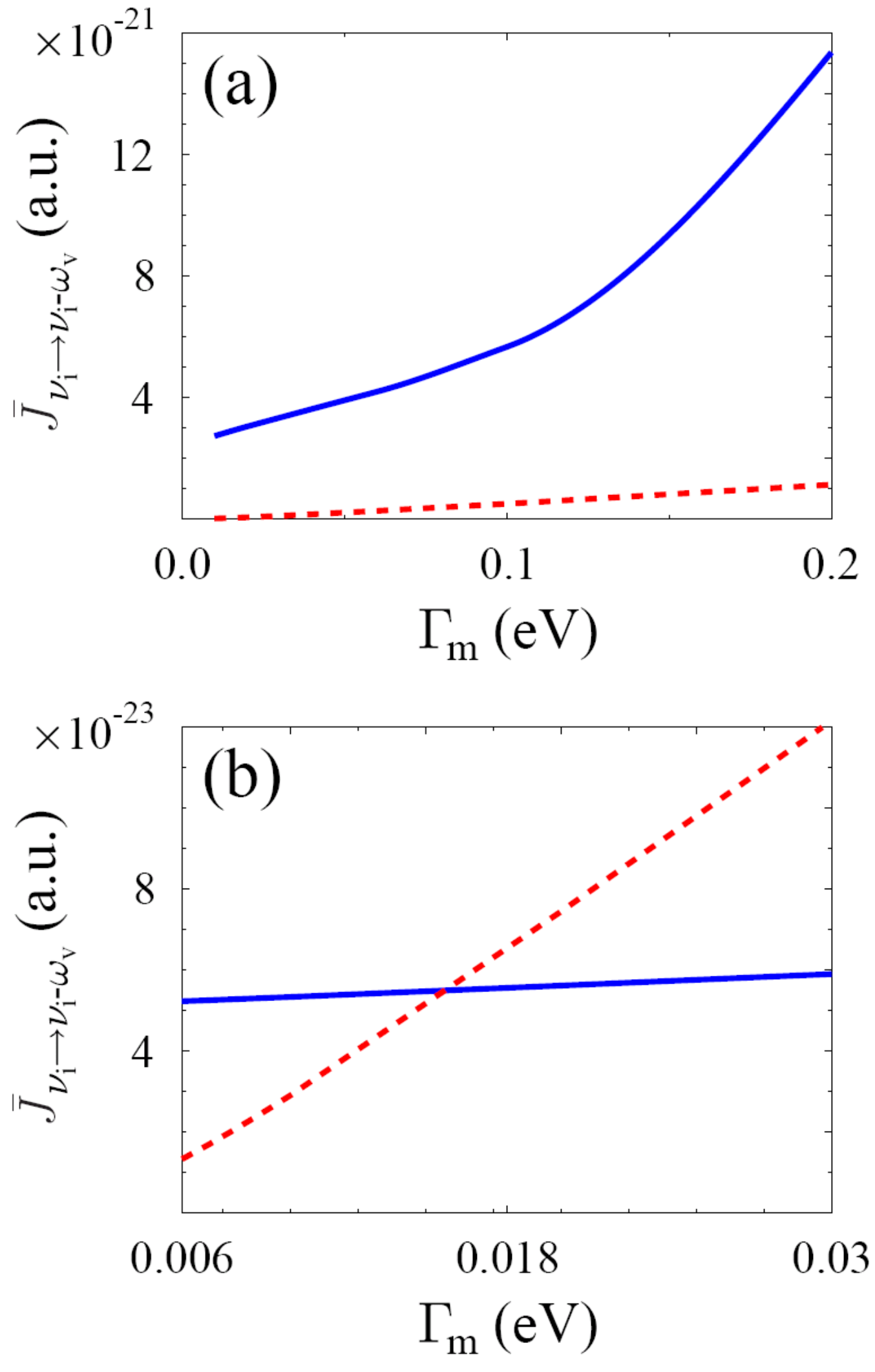

**Figure 7 (Color online)** The Stokes signal, $\nu_f = \nu_i - \omega_\upsilon$, from the normal molecular Raman scattering mechanism, Eq. (56) (solid line, blue) and from the metal-to-molecule Raman process, Eq. (69) (dashed line, red) plotted against the molecule-metal charge-transfer coupling, taken the same for both electrodes and expressed by the magnitude of $\Gamma_m$. Panel (a) shows results obtained for the equilibrium situation, $V = 0$. Panel (b) depicts the equivalent nonequilibrium behavior at $V = 1$V. In both cases the incoming light frequency was chosen to be in resonance with metal-to-molecule transition, $\nu_i = E_2 - \mu$ where $\mu$ is the nearest Fermi-energy, i.e. $\nu_i = 1$eV in (a) and $\nu_i = 0.5$eV in (b). Other parameters are as in Fig.3b.

Finally, we compare, for what we believe to be a reasonable choice of parameters, the contributions from the molecular and the metal-to-molecule transitions to the Stokes Raman signal. Intuitively we expect charge transfer mechanism to be significant in the limit of strong molecule-lead coupling. We find however (Fig.7a) that at equilibrium, the contribution from the charge-transfer mechanism is considerably smaller than that arising from the molecular process for any strength of molecule-metal coupling, even in the resonant charge transfer transition regime.[56] However, the charge transfer contribution may become dominant under applied bias, where the electrochemical potential (Fermi energy) of one of the contacts is closer to molecular level (here the LUMO), as seen in Fig.7b.

## 5. Conclusions

In this paper we have described a theoretical treatment of Raman scattering from biased molecular junctions. We have used a two-level (HOMO-LUMO) model[27, 28] to represent the molecular ground and excited state. These levels are coupled to two metal electrodes and interact with harmonic modes that represent molecular vibrations and the thermal environment, and with the radiation field. The molecule-metal interaction includes both electron and energy transfer terms, the latter accounts for molecular excitation/de-excitation coupled to electron-hole (e-h) generation/destruction in the metals. Two radiative processes are considered: the standard response of an isolated molecule as modified by the metal-lead coupling and light induced metal-molecule electron transfer transitions. Our treatment is based on the non-equilibrium Green function technique, which is used for describing elastic and inelastic photon scattering by the junction as well as the non-equilibrium junction transport. An important difference from our previous studies[27, 28] is the nature of flux under consideration. In [27, 28] we have calculated photon absorption and emission by current-carrying junctions. These are processes of the one-to-all type, where a specified initial state evolves into a continuum of final states. Such processes can be treated rather easily within the NEGF formalism as shown in [27, 28]. Raman scattering, in contrast, is a one-to-one scattering process, where both initial and final states are defined, hence the need to modify the usual NEGF formalism to describe such a process. While the resulting expressions appear general, it should be emphasized that our approximations are valid only close to resonance and any off resonance observations should be regarded as of qualitative nature only.

The resulting expression for the light scattering flux from the model junction contains three additive contributions. The normal scattering component, associated with the ground state of the molecule, is the analog of the equilibrium low temperature process in the isolated molecule and yields the familiar expression for molecular Raman scattering in this limit. The inverse scattering component describes the analogous process associated with the molecule in the excited state. It can become important when the population of this state is significant, as may happen for large enough voltage bias.[44] In the latter case a mixed term resulting from interference between the normal and inverse scattering paths also arises. The integrated and the energy resolved scattering intensities, as well as their normal, inverse and interference components were studied as functions of the bias voltage and the molecule-leads couplings. Noting that the most pronounced effect of the junction environment on the Raman scattering process may be the familiar consequences of the special electromagnetic boundaries conditions encountered in studies of surface enhanced Raman scattering, we have focused in this paper on effects associated with the electrical non-equilibrium in the biased junction. Our main observations were as follows:

(1) The Raman signal depends on the bias voltage, and in particular shows a step change at the conduction threshold. This behavior originates from two different physical effects: The change of molecular level populations when levels enter the window between the leads Fermi energies, and the step-like change in junction temperature when electronic current increases beyond the conduction threshold. The former effect leads to a negative step in all scattering components but it should be kept in mind that in realistic multilevel systems this phenomenon may be masked by scattering processes associated with transitions between other electronic states. The temperature jump beyond the conduction threshold appears as positive step in the anti-Stokes scattering component that can result in an overall positive step in this component.

(2) The interference contribution can add positively or negatively to the main signal that comprises the direct and inverse contributions discussed above. The magnitude of this contribution to the overall signal increases with the electron-hole relaxation rate, $\Gamma^{(e-h)}$, and mostly decreases with increasing molecule-lead electron transfer rate $\Gamma_m$, m=1,2.

(3) The relative intensities of the Stokes and anti-Stokes signals can be used to estimate the temperature in the non-equilibrium junction. Care has to be excercised in taking into account all sources of final frequency dependence of the scattering intensity.

(4) Raman scattering by the molecular and the metal-molecule charge-transfer mechanisms are essentially non-separable. In this paper however we have studied the scattering associated with each of these processes in the absence of the other.

(5) Comparing the independent contributions of the molecular and the charge transfer mechanisms of Raman scattering, we conclude that the molecular mechansim dominates when the incident light is close to the molecular resonance frequency and may or may not dominate the scattering intensity even when the incident radiation is tuned to the energy diffrenece between the molecular level and the metal Fermi-energy. The relative importance of these components depends on the molecule-metal coupling parameters and on the bias potential.

Our theory was compared with previous theoretical studies[9-11] of Raman scattering from molecules adsorbed on metal surfaces by the charge-transfer mechanism. In particular, indications of peak structures in the scattering displayed against the incident frequency or a reference electrode potential were critically examined. We have found that such peaks appear only under strict conditions, in particular only for sufficiently small molecule-metal coupling. Further studies are needed to examine the role of mixed molecular and charge transfer scattering amplitudes.

Raman scattering from molecular junctions is an important new tool in the study of such systems. At the same time, Raman scattering from a non-equilibrium system provides interesting challenges on its own. Future studies call for a better representation of off-resonance Raman scattering from such systems and for a better characterization of the essentially non-separable character of the molecular and the charge-transfer scattering processes. Most important of course is to establish reliable experimental methodologies for studying Raman scattering and other optical processes in the junction environment.

## Appendix A. Derivation of Eq. (20)

The derivation of Eq. (20) follows procedures that were used before to evaluate the electron current[34, 35] and phonon-assisted thermal flux.[36] Here we are interested in the photon flux in a system described by the Hamiltonian (12). This is given by

$$\begin{aligned} J_\alpha(t) &= \frac{d}{dt} < \hat{a}_\alpha^\dagger(t)\hat{a}_\alpha(t) >= i < [\hat{\bar{H}}, \hat{a}_\alpha^\dagger(t)\hat{a}_\alpha(t)] > \\ &= 2Re\left[ iU_\alpha^{(e-p)} < \hat{a}_\alpha(t)\hat{D}^\dagger(t)\hat{X}^\dagger(t) > \right] \\ &\equiv -2Re\left[ U_\alpha^{(e-p)} \mathcal{G}_\alpha^>(t,t') \right]_{t=t'} \end{aligned} \tag{87}$$

where $\mathcal{G}_\alpha^>(t,t')$ is the greater projection of the Keldysh GF

$$\mathcal{G}_\alpha(\tau,\tau') \equiv -i < T_c \hat{a}_\alpha(\tau)\hat{D}^\dagger(\tau')\hat{X}^\dagger(\tau') > \tag{88}$$

The equation of motion for $\mathcal{G}_\alpha(\tau,\tau')$ is

$$\left( i\frac{\partial}{\partial\tau} - \nu_\alpha \right) \mathcal{G}_\alpha(\tau,\tau') = U_\alpha^{(e-p)*} \mathcal{G}(\tau,\tau') \tag{89}$$

where $\mathcal{G}(\tau,\tau')$ is defined in (23). The solution of (89) on the Keldysh contour is

$$\mathcal{G}_\alpha(\tau,\tau') = \int_c d\tau_1 F_\alpha(\tau,\tau_1) U_\alpha^{(e-p)*} \mathcal{G}(\tau_1,\tau') \tag{90}$$

with $F_\alpha(\tau,\tau')$ defined in (22). Taking the greater projection of (90), setting $t = t'$, and substituting the result into (87) leads to (20).

## Appendix B. Derivation of Eqs. (26), (28)-(30)

Here we derive Eqs. (26) and (28)-(30) starting from (25). First we consider the second order corrections to the GF $\mathcal{G}^<(t,t')$ associated with the interaction with the pumping mode $i$. We start by writing $\mathcal{G}^<(t,t')$ explicitly, using the time evolution operator associated with the Hamiltonian (18)

$$\hat{S} = T_c \exp\left[ -i\int_c d\tau \hat{\bar{V}}_I^{(e-p)}(\tau) \right] \tag{91}$$

where $\hat{\bar{V}}_I^{(e-p)}(\tau)$ is operator $\hat{\bar{V}}^{(e-p)}(\tau)$ in the interaction representation with respect to the Hamiltonian (19). We next expand it up to the lowest non-vanishing order (second) in $U^{(e-p)}$. Because the system is non-equilibrium steady state this expansion has to be done on the Keldysh contour. Assuming that only the pumping mode of the radiation field is excited (populated with photons), we get

$$\mathcal{G}^{<}(t,t') = \mathcal{G}_0^{<}(t,t') + \Delta\mathcal{G}^{<}(t,t') \tag{92}$$

$$\begin{aligned}
\Delta\mathcal{G}^{<}(t,t') &= \frac{(-i)^3}{2!}\int_c d\tau_1 \int_c d\tau_2 < T_c \hat{D}^\dagger(t')\hat{X}^\dagger(t')\hat{D}(t)\hat{X}(t) \\
\times\ & [U_i^{(e-p)}\hat{D}^\dagger(\tau_1)\hat{X}^\dagger(\tau_1)\hat{a}_i(\tau_1) + H.c.][U_i^{(e-p)}\hat{D}^\dagger(\tau_2)\hat{X}^\dagger(\tau_2)\hat{a}_i(\tau_2) + H.c.] > \\
&= -|U_i^{(e-p)}|^2 \int_c d\tau_1 \int_c d\tau_2 F_i(\tau_1,\tau_2) \\
&\times < T_c \hat{D}^\dagger(t')\hat{X}^\dagger(t')\hat{D}(t)\hat{X}(t)\hat{D}^\dagger(\tau_1)\hat{X}^\dagger(\tau_1)\hat{D}(\tau_2)\hat{X}(\tau_2) >
\end{aligned} \tag{93}$$

Note that $t$ and $t'$ are regular time variables associated with the time evolution of $\mathcal{G}^{<}(t,t')$ determined by the Hamiltonian $\hat{\bar{H}}_0'$ of Eq. (19), while $\tau_1$ and $\tau_2$ are contour variables. Separating the averaging over field and molecule operators leads to (26).

Next we project the contour variables $\tau_1$ and $\tau_2$, Eq. (26), onto the real time axis. We have to keep in mind two points: 1. The pumping mode $i$ is considered to be a source of photon flux through the system, hence the only projection of $\Pi_i(\tau_1,\tau_2)$ we are interested in is $\Pi_i^{<}(t_1 - t_2)$, i.e. in the steady state we are interested in situation when mode $i$ is occupied and we disregard processes that increase number of quanta in this mode; 2. We are interested only in the diagrams corresponding to rates, i.e. those where interaction with external field connects the upper and lower branches of the contour.[43] In the language of standard diagrammatic technique we retain those time orderings which will contribute to lesser and/or greater self-energies (the ones responsible for in- and out-going fluxes to the system) and disregard contributions to retarded and/or advanced self-energies (the ones describing system damping rates and electronic levels shifts resulting from the process under consideration, in our case due to interaction with external laser field).

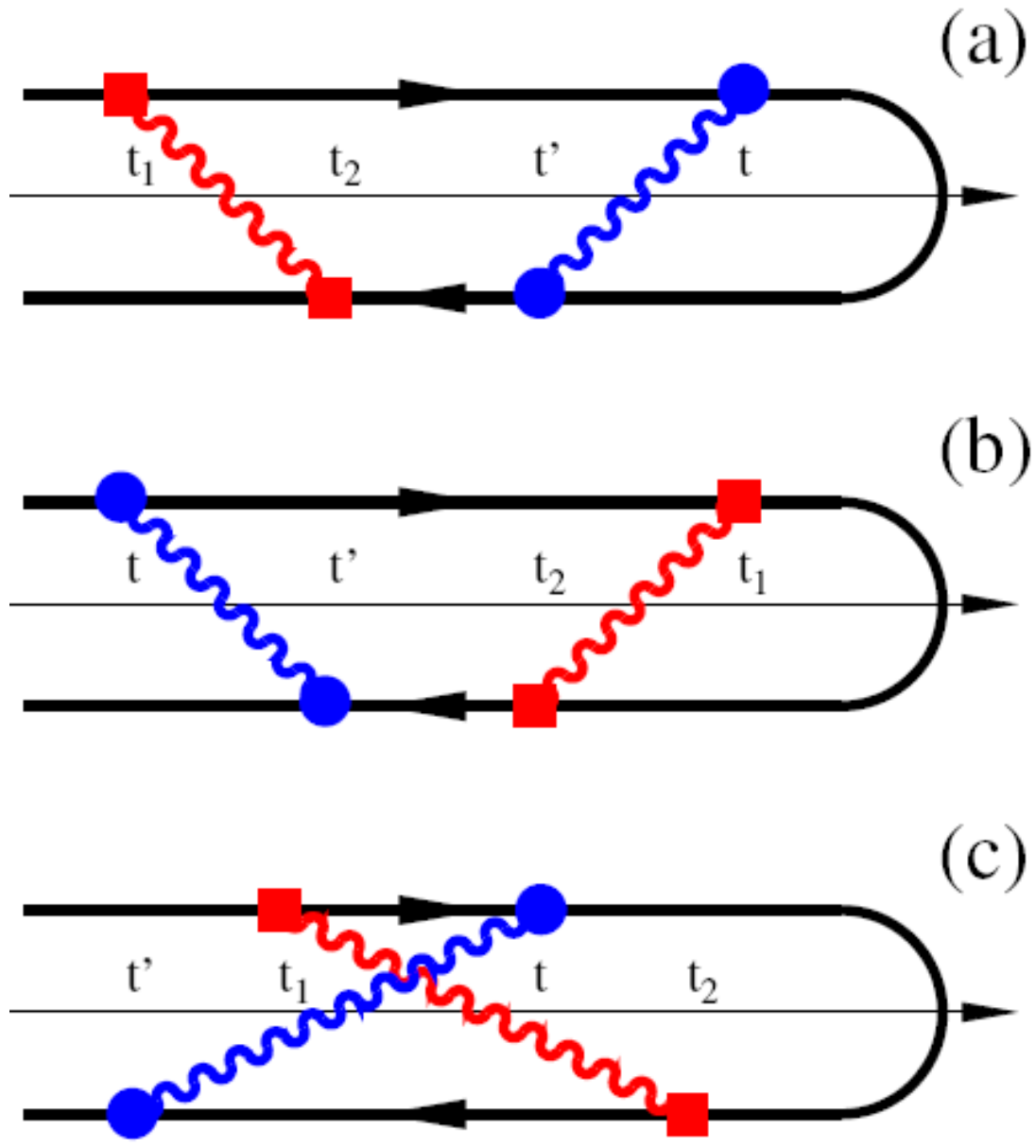

Figure 7: (Color online) Diagrams on the Keldysh contour relevant for calculation of the Raman flux $J_{i\to f}$. Interactions with the source mode $i$ are indicated by squares, while with drain mode $f$ with circles. Wavy lines represent free photon GFs, Eq. (22). Time labels correspond to those in Eqs. (28)-(30). The diagrams correspond to normal Raman scattering (a), inverse Raman scattering (b), and interference between the two (c). Note, there is one more diagram corresponding to an interference process, which is the mirror image (along the vertical axis) of the diagram (c).

With this in mind we are left with diagrams where variables $\tau_1$ and $\tau_2$ are taken at different branches of the Keldysh contour. Moreover, $\tau_1$ should be chosen at the time-ordered branch, while $\tau_2$ at the anti-time-ordered branch in order to guarantee that only inscattering flux from pumping mode $i$ is considered. Finally all possible orderings of $t_1$ and $t_2$ relative to fixed positions of $t$ and $t'$ has to be considered. This leaves us with the diagrams shown in Figure 7. The diagram presented in Figure 7a corresponds to normal Raman scattering, where the initial (and final) electronic state is the ground state of the molecule (or HOMO in single electron language). This can be easily seen from order of say excitation $\hat{D}\hat{X}$ operators along the contour. Indeed, following the contour ordering of Fig. 7a one gets for (93) a contribution of the form

$$\begin{aligned}&-|U_i^{(e-p)}|^2\int_{-\infty}^{+\infty}dt_1\int_{-\infty}^{+\infty}dt_2\,F_i^<(t_1-t_2)\\&\times<T_c\hat{D}(t_2)\hat{X}(t_2)\hat{D}^\dagger(t')\hat{X}^\dagger(t')\hat{D}(t)\hat{X}(t)\hat{D}^\dagger(t_1)\hat{X}^\dagger(t_1)>\end{aligned}\tag{94}$$

The structure of the correlation function in (94) indicates a normal Raman scattering signal. Similarly, Figure 7b represents inverse Raman scattering, when the initial (and final) electronic state is the excited state of the molecule (or LUMO in single electron language). This term vanishes unless the molecular LUMO is populated, which occurs for high enough voltage bias across the junction or at high temperature. Figure 7c is one of the two diagrams (the second is its mirror image along the vertical axis) corresponding to interference between normal and inverse Raman processes.

We use these diagrams together with Eq. (21) and the fact that corresponding free photon GFs for modes $i$ and $f$ are given by

$$\begin{aligned} F_i^{<}(t_1-t_2) &= -ie^{-i\nu_i(t_1-t_2)} \\ F_f^{>}(t'-t) &= -ie^{-i\nu_f(t'-t)} \end{aligned} \tag{95}$$

Then after separation of the GFC factor (correlation function of $\hat{X}$ operators) from the molecular polarization correlation function (correlation function of $\hat{D}$ operators) one gets Eqs. (28)-(30).

## Acknowledgements

The research of AN is supported by the Israel Science Foundation, the US-Israel Binational Science Foundation and by the Germany-Israel Foundation. MG gratefully acknowledges the support of a LANL Director's Postdoctoral Fellowship and of UCSD startup funds. MR thanks the Chemistry Division of the NSF, and the MRSEC program of the NSF, through the Northwestern MRSEC (DMR 0520513), for support.

[56] This result was obtained using the order of magnitude estimate based on Eq. (79), taking $\Gamma_R$ the same molecular radiative rate used to estimate the Raman intensity associated with the molecular process. Note however that $\Gamma_R$ in Eq. (79) is the radiative rate associated with a transition dipole of magnitude (e – electron charge, d – molecule-metal distance)} that may be considerably larger than a typical molecular radiative rate.